\documentclass[prl,aps,
nofootinbib,
floatfix,
superscriptaddress]{revtex4}
\usepackage{graphicx}
\usepackage{epsfig}
\usepackage{rotating}
\usepackage{amssymb}
\usepackage{dsfont}
\usepackage{psfrag}

\newcommand{\beq}{\begin{equation}}
\newcommand{\eeq}{\end{equation}}
\newcommand{\beqs}{\begin{eqnarray}}
\newcommand{\eeqs}{\end{eqnarray}}
\newcommand{\lsim}{\mathrel{\raisebox{-
.6ex}{$\stackrel{\textstyle<}{\sim}$}}}

\renewcommand{\L}{{\cal L}}

\def\hbar{\hspace{0pt}\raisebox{1pt}{$-$} \hspace{-7pt} h}

\def\r{\rho}

\newcommand{\be}{\begin{equation}}
\newcommand{\ee}{\end{equation}}
\newcommand{\bea}{\begin{eqnarray}}
\newcommand{\eea}{\end{eqnarray}}

\def\co{{\cal O}}
\def\lbldef#1#2{\expandafter\gdef\csname #1\endcsname {#2}}

\def\href#1#2{#2}

\newcommand{\ber}{\begin{eqnarray}}
\newcommand{\eer}{\end{eqnarray}}

\newcommand{\beqar}{\begin{eqnarray}}

\newcommand{\eeqar}{\end{eqnarray}}

\newcommand{\dsl}
  {\kern.06em\hbox{\raise.15ex\hbox{$/$}\kern-.56em\hbox{$\partial$}}}

\newcommand{\eeqarr}{\end{eqnarray}}
\newcommand{\ZZ}{{\rm \kern 0.275em Z \kern -0.92em Z}\;}
\def\CC{{\mathchoice
{\rm C\mkern-8mu\vrule height1.45ex depth-.05ex
width.05em\mkern9mu\kern-.05em}
{\rm C\mkern-8mu\vrule height1.45ex depth-.05ex
width.05em\mkern9mu\kern-.05em}
{\rm C\mkern-8mu\vrule height1ex depth-.07ex
width.035em\mkern9mu\kern-.035em}
{\rm C\mkern-8mu\vrule height.65ex depth-.1ex
width.025em\mkern8mu\kern-.025em}}}

\def\RR{{\rm I\kern-1.6pt {\rm R}}}

\def\ZZ{{\rm Z}\kern-3.8pt {\rm Z} \kern2pt}
\def\IB{\relax{\rm I\kern-.18em B}}
\def\ID{\relax{\rm I\kern-.18em D}}
\def\II{\relax{\rm I\kern-.18em I}}
\def\IP{\relax{\rm I\kern-.18em P}}

\newcommand{\rc}{\nonumber\\}
\newcommand{\bear}{\begin{eqnarray}}
\newcommand{\eear}{\end{eqnarray}}

\newcommand{\x}{{\cal O}}

\def\to{\rightarrow}

\def\to{\rightarrow}

\def\a{\alpha}

\def\f{\phi}               %      \varphi
\def\vf{\varphi}

\def\l{\lambda}
\def\m{\mu}
\def\n{\nu}
  \def\w{\omega}
  \def\th{\theta}                  %     \vartheta
\def\r{\rho}                                     %     \varrho
\def\t{\tau}

\def\x{\xi}
\def\z{\zeta}

\def\lab{\label}
\def\6{\partial}
\def\wg{\wedge}

\def\NO{\nonumber}

\def\bea{\begin{eqnarray}}
\def\eea{\end{eqnarray}}

\def\beqx{\begin{displaymath}}
\def\eeqx{\end{displaymath}}

\newcommand{\bmat}{\left(\begin{array}}
\newcommand{\emat}{\end{array}\right)}

\def\a{\alpha}

\def\f{\phi}

\def\l{\lambda}
\def\m{\mu}
\def\n{\nu}

    \def\om{\omega}

    \def\th{\theta}
\def\r{\rho}

\def\t{\tau}
\def\x{\xi}
\def\z{\zeta}

\def\L{\Lambda}

    \def\Om{\Omega}

\def\vf{\varphi}

\def\co{{\cal O}}

\def\bo{{\raise-.3ex\hbox{\large$\Box$}}}               % D'Alembertian
\def\pa{\partial}                                       % curly d
\def\face{{\raise.2ex\hbox{$\displaystyle \bigodot$}\mskip-2.2mu \llap {$\ddot
        \smile$}}}                                   % happy face
\def\>{\rangle}                                      %right angle
\def\<{\langle}                                      %left angle

\newcommand{\sub}[1]{\phantom{}_{(#1)}\phantom{}}    % subscript in ( )
\def\leftrightarrowfill{$\mathsurround=0pt \mathord\leftarrow \mkern-6mu
        \cleaders\hbox{$\mkern-2mu \mathord- \mkern-2mu$}\hfill
        \mkern-6mu \mathord\rightarrow$}        % <--> double differential
\def\dvec#1{\vbox{\ialign{##\crcr
        \leftrightarrowfill\crcr\noalign{\kern-1pt\nointerlineskip}
        $\hfil\displaystyle{#1}\hfil$\crcr}}}           % <--> accent
\def\-{\hphantom{-}}

\begin{document}
\title{Walking Dynamics from String Duals
%RG flow with walking dynamics from gravity duals. **TO IMPROVE**
}

\author{Carlos N\'u\~nez, Ioannis Papadimitriou and Maurizio Piai}
%\author{Ioannis Papadimitriou}
%\author{Maurizio Piai %\thanks{email:piai@u.washington.edu }
%\thanks{email:M.Piai@Swansea.ac.uk}
%\affiliation{Department of Physics, University of Washington, Seattle, WA 98195}
%}
\affiliation{Swansea University, School of Physical Sciences,
Singleton Park, Swansea, Wales, UK}

\date{\today}

%\vspace{6mm}

\begin{abstract}
Within the context of a string-theory dual to ${\cal N} = 1$ gauge
theories with gauge group $SU(N_c)$ and large $N_c$,
we identify a class of solutions of the background equations
for which a suitably defined dual of the gauge coupling exhibits
the features of a walking theory. We find evidence for three distinct, dynamically generated scales,
characterizing walking, symmetry breaking and confinement, and we put them
in correspondence with field theory
by an analysis of the operators driving the flow.
%We follow  the running of the gauge coupling from the UV,
%where the model is weakly coupled.
%At some high scale the running becomes virtually flat,
%entering a regime in which the running is suppressed (walking).
%At a second, lower scale, the walking behavior stops,
%and the more familiar logarithmic running of the coupling appears.
%At a third, even lower scale, a (good) singularity develops, signaling
%the break-down of the description in terms of weakly coupled UV degrees of freedom,
%and the formation of bound states (confinement).
%A set of constraints on the integration constants,
%derived by avoiding singular behaviors in extrapolating the  solution
%to the far UV and far IR, is
%interpreted in terms of the naturalness of the field theory.
%If these constraints are imposed,
% the geometry has no singularity (aside from the one at confinement),
% and hence the corresponding RG evolution
%of the couplings in the dual field theory
%is well  behaved up to asymptotically
%high energies.
%We also analyze some of the features of the dual field theory.
%The field content of the theory is not known,
%but we identify the scaling dimensions of the operators and
%condensate that act as relevant deformations,
%by analyzing the UV expansion of the
%background.
%We also discuss
%the structure of approximate
%symmetries of the model, by studying the approximate isometries of the
%background (metric, dilaton and Ramond-Ramond 3-form).
\end{abstract}

%\pacs{11.10.Kk, 12.15.Lk, 12.60.Nz}

\maketitle

\tableofcontents

%%%%%%%%%%%%%%%%%%%%%%%%%%%%%%%%
%%%%%%%%%%%%%%%%%%%%%%%%%%%%%%%%
\newpage
\section{Introduction and general ideas}

In this paper we propose to study the  features of walking dynamics
within the context of the string theory construction of
background geometries which are conjectured to be
a dual description of strongly coupled four-dimensional systems.
We dispense with the model-building details of a complete technicolor model,
in particular with electro-weak symmetry-breaking itself.
Instead, we focus completely on the properties of walking dynamics in isolation.
We find an explicit solution to the
supergravity equations yielding
a background which admits an interpretation in terms of
a gauge theory, the coupling of which exhibits the qualitative behavior of
a walking model, and study the properties of this
solution.

The advantages of following this program are multiple.
First of all, there is a set of well-defined and controllable expansions allowing
for a systematic calculation to be performed.
The running of the gauge coupling is defined in terms of the geometry.
It can be tracked all the way into the strongly coupled region
where traditional perturbative techniques cannot be trusted.
As a result, it is possible to identify
the transition in the far UV below which the running flattens.
It is also possible to follow the dynamics in the far IR, where
due to the appearance of a non-trivial condensate
 (in these specific models usually identified with the gaugino condensate),
 which breaks spontaneously some of the global symmetry,
the running reappears, ultimately leading to confinement.
It is possible to show that several separate scales are
 dynamically generated, without introducing UV-singularities (fine-tuning)
 in the background.
Interestingly, we find that the existence of this background is completely
independent of the
presence of flavors, indicating that its dynamical origin does not necessarily arise from
the interplay of $N_c$ and $N_f$ in the beta-functions.
This suggests that walking dynamics does not necessarily
require the presence of large
numbers of fundamental fermions, which is a very welcome
feature for model-building.

This is the first step in a relatively unchartered territory.
In principle, besides studying the RG evolution of the underlying model,
which is the main focus of this paper,
it should be possible to determine all the symmetry
and symmetry-breaking pattern of the model,
its low-energy spectrum of composite states, including vector mesons,
dilaton and pseudo-Nambu-Goldstone
bosons (PNGBs),
together with the field content of the dual gauge theory and
all the anomalous dimensions in the IR.
In practice, this is a quite extensive and challenging  research program
which will be completed elsewhere.

The paper is structured as follows.
We start by reviewing the motivations for walking dynamics
and the open questions that arise in this context.
This section is intended for the reader who is not familiar with
dynamical electro-weak symmetry breaking.
We then review the basics of the string-theory construction of a particular
class of models that are believed to be dual to a broad class of
strongly coupled gauge theories, we write the equations
that are going to be used in the body of the paper, and we fix the notation.
This is mostly a summary of previous results that can be found in the literature,
and is intended for the reader who is not acquainted with this framework.
We then introduce a new class of solutions to the background equations,
and develop the study of the properties of this solution in the body of the paper.
We construct the solution via a systematic expansion, discuss
its  properties, perform an analysis of scaling dimension of the operators
deforming the dual field theory, and of the symmetries of the resulting background.
In particular, we show that the running of a suitably defined
four-dimensional gauge coupling has, within this class of solutions, the
basic properties expected in a walking theory.
We conclude by summarizing a tentative research program based on these results.

%%%%%%%%%%%%%%%%%%%%%%%%%%%%%%%%
%\newpage
\section{Aspects of Walking Technicolor}

\subsection{The problems of QCD-like technicolor models}

Spontaneous electro-weak symmetry breaking might be induced
by the dynamical formation of a condensate in a new (strongly-coupled) sector
of the complete Lagrangian extending the Standard Model.
In its original form~\cite{TC},
this idea is implemented by assuming that a new non-abelian technicolor
gauge symmetry be present, typically $SU(N_c)$, with a generic
number of technicolors $N_c$,
 that a new set of $N_f$ fermions transforming according to the
fundamental representation of $SU(N_c)$ are present,
and that a $SU(2)\times U(1)$ subgroup of the global symmetry group of the new
technicolor sector be (weakly) gauged in order to reproduce the electro-weak
gauge group of the Standard Model.
If the dynamics of the technicolor sector is
similar to the one of QCD (in which case we refer to this scenario as QCD-like technicolor),
at high scales the new interaction is asymptotically free,
but quantum effects dynamically produce a physical scale
$\Lambda_{TC}$ at which the coupling becomes strong,
the theory confines, and the global (chiral) symmetry is spontaneously broken by the
formation of a condensate of techniquarks. If $\Lambda_{TC}$ is the electro-weak scale,
the induced breaking of $SU(2)\times U(1)$ would produce the physical masses of the
$W$ and $Z$ gauge bosons.

This idea is particularly appealing because it provides a completely natural solution to the
hierarchy problem.
Three major, correlated difficulties arise
when trying to build a realistic model implementing this idea.
The first obstacle is that, by definition, such a scenario
requires the model to be strongly coupled at the electro-weak scale,
and hence the phenomenology of electro-weak interactions
cannot be analyzed with standard perturbative techniques.
One completely model-independent way of dealing with this relies on
the idea of constructing a  low-energy
effective field theory (EFT) description of the interactions
of the electro-weak gauge bosons, hence encoding in the
coefficients of the electro-weak chiral Lagrangian~\cite{EWCL}
all the  information about the (strong) technicolor dynamics.
While this approach is systematic and elegant,
it has a somewhat limited predictive power,
in particular because it treats the precision electro-weak parameters
(such as the  $T$ and $S$ parameters~\cite{PT}) as free parameters to be
extracted from the data, rather than deriving them from first principles.
It is only possible, within this approach, to construct arguments
that yield an order-of-magnitude estimate of the expected size
of the coefficients, based on the power-counting that arises from the
perturbative expansion of the chiral Lagrangian itself. We call this naive
dimensional analysis
(NDA)~\cite{NDA}.

In QCD, the analog construction yields an EFT in which the NDA expectations are in substantial
agreement both with the experimental data and with first principle lattice calculations.
It is hence reasonable to assume that, provided the underlying dynamics be QCD-like,
the  NDA expectations, corrected by appropriately rescaling (in energy and in $N_c$)
the established results of QCD, should give an acceptable prediction for the
 coefficients of the electro-weak chiral Lagrangian, and in particular for the precision parameters.
 However, this is in sharp contrast with the results from
 the combined fit of the experimental data on precision electro-weak physics~\cite{Barbieri},
which seem to indicate that the upper limits on the precision parameters are
quite tight, generically one order of magnitude below the NDA expectations.
Unless one speculates that the $N_c$ and $N_f$ scalings of the QCD-like theory can be
extrapolated all the way into the small-$N_c$ region, and is hence
going to conclude that $N_c=2$
is allowed by the data.

By itself, the fact that NDA estimates are in excess of the experimental data
 might just indicate that the expansion parameter of the
electro-weak chiral Lagrangian be somewhat smaller than expected on the basis of
NDA, which by itself might even be seen as
a welcome, though unexpected and puzzling, feature.
But a third difficulty arises when considering the generation of the
standard-model masses and mixing. In order to understand how this arises,
one has to  remember that in the absence of a  Higgs,
the standard model masses are introduced via extended technicolor (ETC)\cite{ETC}.
Effectively, ETC consists of adding to the Lagrangian a set of irrelevant operators coupling
the standard-model fermions to the new technifermions, in such a way that
the formation of technifermion condensates produces, via dimensional transmutation,
a low-energy description in which mass terms for the standard-model fermions
are generated. This has to be done in such a way as to avoid
introducing significant new sources of flavor changing neutral current (FCNC) interactions,
which typically requires to assume that some family symmetry be broken (sequentially)
at scales $\Lambda_{ETC} \gg \Lambda_{TC}$.
The resulting mass is going to be proportional to the electro-weak scale,
via a coefficient that results from the matching of the appropriate higher-order operator,
which typically is a dimension-6 four-quark  interaction,
with coefficient ${\cal O}(1/\Lambda_{ETC})^2$,
onto the mass term, and is hence going to be suppressed by $\Lambda^2_{TC}/\Lambda_{ETC}^2$.
The fact that the mass of the top quark be as large as the electro-weak
scale itself is hard to reconcile with this scenario, and introduces a further
tension on power-counting arguments within the electro-weak chiral Lagrangian.

To summarize: in the absence of a systematic calculational tool,
the predictive power of a technicolor model is very limited,
and relies on NDA estimates of the coefficients of the low-energy EFT
that are supported only by the experience with QCD.
On the one hand, precision electro-weak data can be reconciled with the
EFT  treatment  only at the price of assuming that NDA provides a systematic {\it overestimate}
of the effects of dynamical electro-weak
symmetry breaking (DEWSB)  on the  electro-weak gauge bosons. On the other hand
the measured large top mass can be reconciled with the
suppression of  flavor changing neutral current (FCNC) interactions only
at the price of assuming that NDA {\it underestimates}
the effects of DEWSB  in the generation of the fermion masses.
All of this constitutes strong, though not definitive, arguments disfavoring this scenario.

\subsection{The solutions provided by walking technicolor}

The line of thinking leading to this (premature) conclusion is, however, flawed at its core:
the NDA estimates of the coefficients of the EFT describing a strongly-coupled
system are based on non-rigorous arguments, that yield acceptable results for QCD,
or for QCD-like theories, only.
In particular, QCD-like theories are characterized by having only one
dynamical scale, and by  the fact that the spectrum of anomalous dimensions
of the theory is, almost at all scales, perturbative, due to the specific
shape of the QCD renormalization group (RG) flow.
One can think of justifying  the NDA  counting rules yielding the
results summarized so far with arguments that rely on these two properties
of QCD.
But if DEWSB results from a model which is not QCD-like, none of these arguments
is justified. This is the case for walking technicolor~\cite{WTC}.

Walking dynamics is an essential ingredient in the modern construction
of models of DEWSB.
The basic assumption is that the condensate inducing electro-weak symmetry breaking
emerges from a strongly coupled sector that, rather than being QCD-like,
is quasi conformal and strongly coupled
 over  a significant range of energy above the electro-weak scale.
This naturally leads to the coexistence of
parametrically separated dynamical scales, and to the appearance of
large anomalous dimensions with important phenomenological implications.
As a result, at the EFT level,
very large departures from the estimates of NDA
cannot be excluded, but are a natural expectation.
The experimental results on electro-weak precision parameters,
flavor-changing neutral currents
and quark masses can hence be reconciled with this framework.
The assumption of walking allows to
construct realistic models that are testable at the LHC~\cite{APS}.

To be more specific, let us remind the reader about what are the
properties of a putative walking technicolor model.
While a QCD-like model is characterized
by only one dynamically generated scale, the structure of a walking
theory implies the existence of four distinct dynamical scales.
A generic model might be  asymptotically free in the far UV,
because the RG flow is assumed to have a trivial UV fixed point.
Due to the fact that the gauge coupling is marginally relevant,
following the RG flow to lower energies the gauge coupling grows,
until a first dynamical scale $\Lambda_{*}$ appears.
The RG equations are assumed to posses a (approximate) fixed point in the IR
at strong coupling.
Below $\Lambda_{*}$ the running of the gauge coupling is almost flat,
because the theory, while approaching its IR fixed point, is approximately conformal.
The fact that the IR fixed point is strongly coupled implies that the spectrum of anomalous dimensions of the operators in the theory is expected to be radically different from the
perturbative results obtained in proximity of the trivial UV fixed point.
Examples of this dynamical behavior exist, for instance  in the well understood context of
supersymmetric QCD ~\cite{seiberg}\cite{cascade}.

The fixed point is only approximate, and the flow, after spending some time ({\it walking})
in its vicinity, will drift away from it. At this point, the approximate scale invariance is broken.
Below this scale $\Lambda_{IR}$ the gauge coupling will start running again,
and ultimately become big enough to induce confinement at a scale $\Lambda_0 < \Lambda_{IR}$.
The condensate that spontaneously breaks the global symmetries of the model
must also form at the electro-weak scale $\Lambda_{TC}$,
in order to induce DEWSB. However, while in QCD
this condensate arises at the confinement scale (temperature),
in general its formation
might  take place at a higher scale (temperature), in the range
$\Lambda_0 < \Lambda_{TC} < \Lambda_{IR}$.
While it is hard to believe that
these three scales (temperatures) can be separated by arbitrarily large factors,
the idea that chiral symmetry might break at a scale (temperature)
somewhat higher than the confinement scale
has been discussed in the past~\cite{CC},
and recently been revived both in the context of lattice studies~\cite{lattice}
and of string-theory inspired models~\cite{SS}.

If this is the underlying dynamics, the NDA estimates of the precision parameters
can very plausibly be modified in a quite substantial way.
An early study in this direction~\cite{AS}  highlighted that
if the anomalous dimension of the techniquarks is large enough,
so that the chiral condensate is effectively dimension-2,
the Weinberg sum rules are going to be modified, and as a consequence
the predictions of the precision parameter $S$ based on dispersion relations
must be revised.
This fact, together with the fact that several distinct scales are dynamically generated
in the UV, has been the subject of several recent EFT studies inspired by the
ideas of the AdS/CFT correspondence~\cite{AdS/TC},
leading to the conclusion that in this context large regions of the parameter space
of the models are compatible with the precision electro-weak studies and
are potentially testable at the LHC.
At the same time, the large anomalous dimension of the chiral condensate
might also provide a natural solution to the problem of the mass of the top,
without introducing new significant sources of FCNC~\cite{APS}.

\subsection{Open questions from walking dynamics}

The conclusion is that if DEWSB is realized in nature,
walking  is very likely to be its crucial dynamical feature.
But the fact that the dynamics be strongly coupled, and so different from
QCD, is a major obstacle for the search of viable candidate
theories in which walking emerges as a dynamical feature,
rather than being put in by hand as a working assumption in the EFT.

Very little is known about the implications
of walking dynamics by itself.
There are a set of well-defined field theory questions
which are of utmost importance from the phenomenological
perspective and which cannot be addressed within the low-energy EFT,
but require to have a dynamical model.
For the most part, these are questions that are
independent of the details of how the strongly coupled sector
is coupled to the standard model fermions and gauge bosons,
and of electro-weak symmetry breaking.
In view of this, it is very useful to have a complete model
in which walking emerges, such that the dynamical implications of walking can be
studied in isolation, factoring if out for the complicated structure
of a complete model of dynamical electro-weak symmetry breaking.

One set of such questions has  been anticipated in the previous discussion,
and  includes the identification of the (four) dynamical scales in the system,
the study of the mechanism leading to their formation
and their separation.
We also anticipated the importance of
 calculating of the spectrum of anomalous dimensions
of the underlying theory
in the walking regime.

Besides these, an even more fundamental
set of questions is related to the study of the approximate global
symmetries of the model. If the underlying dynamics is approximately conformal,
(or if it possesses approximate internal global symmetries
not related to the electro-weak symmetry)
and the formation of condensates breaks the dilatation invariance
(or any of the global internal
 symmetries) spontaneously,
in principle a light dilaton (or a set of light pseudo-Nambu-Goldstone bosons)
might be present in the model.
The presence of any light scalar might  change in a radical
way the low-energy phenomenology, and it is important to understand
under what dynamical conditions they arise.
In particular, because of the dilaton quantum numbers, and of the quantum numbers of the Standard Model Higgs, distinguishing the two at a hadronic machine such as the LHC is
going to be a major challenge.
It is hence important to know if walking dynamics predicts the existence
of such a dilaton, and what its mass and couplings  are going to be.
This has been investigated for a long time, but is still is a very open problem~\cite{dilaton}.

%%%%%%%%%%%%%%%%%%%%%%%%%%%%%%%%
%         Addendum to walking string  
%%%%%%%%%%%%%%%%%%%%%%%%%%%%%%%%

\subsection{Going for a walk.}

The potential of walking dynamics did not go unnoticed. 
A variety of studies exists in the literature, within the context
of four-dimensional non-abelian gauge theories,
looking for models in which a (approximate) fixed point
of the RG equations exists in the IR, as sensible candidate for
a walking technicolor model.

This is a very challenging problem for analytical calculations.
Within the regime in which perturbation theory can be fully trusted,
it has been established long ago~\cite{BZ} that
due to the different dependence on $N_c$ and $N_f$ of the 1-loop and
2-loop beta function, such fixed points exist for certain regimes of $N_f/N_c$, 
a fact that has been studied systematically  
in the context of  supersymmetric theories~\cite{seiberg}.
Studies based on approximate techniques 
attempted to generalize this beyond the regime of perturbation theory
for non-supersymmetric theories,
and seem to indicate that there exist models in which
strongly coupled fixed points and large anomalous dimensions emerge dynamically~\cite{KS}.
More recently, similar results have been obtained by analyzing the RG flow
with the use of of a conjectured generalization 
of the NSVZ~\cite{NSVZ} beta-function 
to non-supersymmetric set-ups~\cite{exactbeta}.
However, besides being based on ad hoc approximations that require independent 
testing, none of these studies  deal quantitatively with the more realistic
scenario in which the fixed point is only approximate, and hence, after
a period of walking, the theory ultimately confines.

In very recent years, progress on the lattice allowed to perform  numerical 
studies looking for some evidence of the existence of fixed points in the IR,
exact or approximate, in a variety of models.
One such study~\cite{AF} seems to confirm that  fixed points 
in the spirit of~\cite{BZ} exist even beyond perturbation theory.
More elaborate studies confirm this result also for models
with matter in higher-dimensional representation~\cite{DSS}.
A number of collaborations has been testing these same ideas with
complementary techniques~\cite{walkthelattice}.
However, also in this case it is not yet possible to discuss in quantitative
detail the full set of transitions that makes the models first enter the walking regime, then
(at lower energies) leave it, and ultimately confine.

In this paper we propose an alternative approach, based on  gauge/string correspondence,
within which to carry on the program of looking for models that exhibit 
the dynamical features of a walking theory.  
This approach might help shedding some light over
model-independent phenomenological and theoretical features of 
a large class of models with walking dynamics.
 Within this context, it should be possible to address
the set of well posed theoretical questions summarized in the previous subsections,
and the results might be helpful even in guiding the data analysis of 
studies performed using more traditional approaches.

%%%%%%%%%%%%%%%%%%%%%%%%%%%%%%%%%%%%%%
%%%%%%%%%%%%%%%%%%%%%%%%%%%%%%%%%%%%%%
%\newpage
\section{Aspects of the String model and the dual QFT}\label{themodel}
%{\it The discussion in this section is targeted mostly to people not working
%on String Theory.}

In this section we will specify the type of string duals to strongly
coupled field theories that we will be studying.

There are various ways of constructing string duals to four dimensional
theories. We will focus on the models that use wrapped branes. To make
things
concrete, suppose that we consider at first $N_c$ D5 branes. The dynamics of
these
five-branes is well described at low energies by a $(5+1)$ field theory
with
sixteen SUSY's. The Lagrangian is the dimensional reduction of
ten dimensional Super-Yang-Mills to six dimensions. The field content can
be seen to be: a gauge field $A_{M}$, two sets of spinors
$\l_l,\;\tilde{\l}_{\bar{l}}$ (with four
components each) and four real scalars $\phi_i$, all in the adjoint
representation of $SU(N_c)$. The presence of the branes
breaks the $SO(1,9)$
symmetry group of Type IIB supergravity into $SO(1,5) \times
SO(4)\approx SU(4)\times SU(2)_A \times SU(2)_B$.
We summarize the field content and transformation laws in
Table~\ref{Table:fields}.

\begin{table}
\begin{center}
\begin{tabular}{l c c c}
\hline
& $SU(4)$ & $SU(2)_A$ & $SU(2)_B$ \\
\hline
$A_M$ & ${\bf 6}$ & ${\bf 1}$ & ${\bf 1}$ \\
$\phi_i$ & ${\bf 1}$ & ${\bf 2}$ & ${\bf 2}$ \\
$\lambda_l$ & ${\bf 4}$ & ${\bf 2}$ & ${\bf 1}$\\
$\tilde{\lambda}_{\bar{l}}$ & ${\bf \bar{4}}$ & ${\bf 1}$ & ${\bf 2}$ \\
\hline
\end{tabular}
\end{center}
\caption{The six-dimensional field content of the $D5$ system and their
transformation under global symmetries.
\label{Table:fields}}
\end{table}

The dual description of this field theory at strong coupling is given by the background generated by $N_c$
D5 branes after the decoupling limit
\beq
g_s\to \infty,\;\;\; \alpha' \to 0,\;\;\;\; N_c\to\infty\;\;\; g_s\alpha'
N_c=fixed
\eeq
 is taken \cite{Itzhaki:1998dd}. This
background consists of a metric, a RR three form and a dilaton and reads,
\be\lab{D5sol}
ds^2 = e^{\f}\left[dx_{1,5}^2 +\a'g_s
N_c(dr^2+\frac{1}{4}\sum_{i=1}^3 \tilde{\om}_i^2)\right],\; F_{(3)}  =
\frac{N_c}{4}
  \tilde{\w}_1 \wg \tilde{\w}_2\wg \tilde{\w}_3,\; e^{\f}=e^{\phi_0+ r} .
\ee
We define the $SU(2)$ left-invariant one forms as,
\bea\lab{su2}
\tilde{\w}_1&=& \cos\psi d\tilde\theta\,+\,\sin\psi\sin\tilde\theta
d\tilde\varphi\,\,,\rc\rc
\tilde{\w}_2&=&-\sin\psi d\tilde\theta\,+\,\cos\psi\sin\tilde\theta
d\tilde\varphi\,\,,\rc\rc
\tilde{\w}_3&=&d\psi\,+\,\cos\tilde\theta d\tilde\varphi\,\,.
\eea
Hence,
\beq
\sum_{i=1}^3 \tilde{\w}_i^2=d\tilde{\theta}^2
+\sin^2\tilde{\theta}d\tilde{\varphi}^2
+ (d\psi+\cos\tilde{\theta} d\tilde{\varphi})^2,
\eeq
is a nice compact way or writing a three-sphere. The ranges of the three
angles
are $0\le\tilde\varphi< 2\pi$,
$0\le\tilde\theta\le \pi$, $0\le\psi< 4\pi$.

Now, to produce a four dimensional effective theory out of the previous six
dimensional one, we can imagine separating two directions, say $(x_4,
x_5)$,
and wrapping these $N_c$ D5 branes on a small
two-sphere (in other words, compactifying the $(x_4,x_5)$
space),
so that low energy modes will explore only the
non-compact $3+1$ directions. There are different ways of choosing the
two-dimensional space. All of them will lead to field theories describing
the
excitations of the wrapped five branes. For different technical reasons it
is convenient to choose a two-dimensional space that preserves some
fraction
of the 16 SUSY's. (For example, a torus preserves the sixteen
supercharges.) We will be interested in effective 4-d theories that
preserve a
minimal amount of SUSY (four supercharges in 4-d). This example was worked
out in \cite{Maldacena:2000yy}. (See also the paper \cite{Andrews:2006aw}
for interesting details.)
One can see that this system preserves (or
partially breaks) a fraction of the original SUSY via a quite general
`twisting procedure' nicely explained by Witten in \cite{Witten:1994ev}.
One effect of this `unconventional' way of breaking SUSY is that even at
high energies, the background preserves only four supercharges, as can be
seen, by studying the weakly coupled spectrum
\cite{Andrews:2006aw}.

Let us be more precise about the twisting procedure. If we take the fields
for the $N_c$ flat D5 branes, they transform as explained under
$SO(1,5)\times SO(4)$, see
the Table I  above. Now, when we separate two directions, for
example $(x_4,x_5)$
and compactify them on a two-sphere $S^2$, we are breaking the global
transformation group into
\beq
SU(4)\times SU(2)_A \times SU(2)_B \to SU(2)_L
\times SU(2)_R \times U(1)_{45}
\times  SU(2)_A \times SU(2)_B
\label{branching}
\eeq
where we made $SO(1,3)\approx SO(4)=SU(2)_L
\times SU(2)_R$. Now, we need to decompose
the fields
under the `branching' described in eq.(\ref{branching}). This was
carefully done in Section 3 of the paper \cite{Andrews:2006aw}. The
next step is the `twisting' (or mixing) between the global symmetries
of the $S^2$ and a $U(1)$ factor inside $SU(2)_A$, which produces a set of 4-d
fields with or without transformation rule under the new combined $U(1)$.
This twisting procedure allows branes wrapping
cycles to preserve some amount of SUSY, in spite of the cycle
not necessarily admitting massless spinors \cite{Bershadsky:1995qy}.

The `twisted-Kaluza-Klein' decomposition is explained in detail in
Section 4 of \cite{Andrews:2006aw}. We will just need to state that the
spectrum -at weak coupling- consists of a massless vector multiplet
$W_\alpha$ whose action is the usual N=1 SYM one and (after a mass gap
related
to the inverse size of the sphere) an infinite tower of massive chiral
multiplets and massive vector multiplets. The degeneracies and masses are
given in \cite{Andrews:2006aw} where the field theory was proven to be
equivalent to $N=1^*$ SYM when expanded around a particular Higgs vacuum.
Also, the fact that there is an infinite tower of
states reminds us that the field theory is higher dimensional at high
energies.
Let us stress that these are all {\it{weakly coupled results}}.

The natural question is if there is a non-perturbative description of the
field theory living on the wrapped five branes described in the previous
paragraph. This is indeed the case.
One has to construct a background that represents the physical situation
described above, by compactifying on a sphere and twisting the background
of eq.(\ref{D5sol}). The background (in string-frame) reads
\bea
ds^2 &=& \alpha' g_s e^{ \phi(\rho)} \Big[\frac{dx_{1,3}^2}{\alpha' g_s} +
e^{2k(\rho)}d\rho^2
+ e^{2 h(\rho)}
(d\theta^2 + \sin^2\theta d\varphi^2) +\nonumber\\
&+&\frac{e^{2 g(\rho)}}{4}
\left((\tilde{\omega}_1+a(\rho)d\theta)^2
+ (\tilde{\omega}_2-a(\rho)\sin\theta d\varphi)^2\right)
 + \frac{e^{2 k(\rho)}}{4}
(\tilde{\omega}_3 + \cos\theta d\varphi)^2\Big], \nonumber\\
F_{3} &=&\frac{N_c}{4}\Bigg[-(\tilde{\omega}_1+b(\rho) d\theta)\wedge
(\tilde{\omega}_2-b(\rho) \sin\theta d\varphi)\wedge
(\tilde{\omega}_3 + \cos\theta d\varphi)+\nonumber\\
& & b'd\rho \wedge (-d\theta \wedge \tilde{\omega}_1  +
\sin\theta d\varphi
\wedge
\tilde{\omega}_2) + (1-b(\rho)^2) \sin\theta d\theta\wedge d\varphi \wedge
\tilde{\omega}_3\Bigg].
\label{nonabmetric424}
\eea
In the following, we will set units so that $\alpha'=g_s=1$.
Notice that the following `deformations' with respect to the `flat'
five-brane
metric in eq.(\ref{D5sol}) have been implemented,

i) Two of the directions in $R^{1,5}$ have been renamed as
$(\theta,\varphi)$
and we have compactified them on a  two-sphere.

ii) There is a {\it mixing}  (fibration) between the $(\tilde{\theta},\tilde{\varphi},
\psi)$
coordinates and the $(\theta,\varphi)$ ones. This mixing is encoded by the
presence of  the
functions $a(r), b(r)$ and in the fact that the last component of the metric
is $(d\psi +\cos\tilde{\theta} d\tilde{\varphi} +\cos\theta d\varphi)$.

iii) For a given set of conditions on the functions (see below), the
background in eq.(\ref{nonabmetric424}) was shown to preserve four
supercharges, hence being dual to a  four dimensional $N=1$ theory.

iv) The functions in the background above must solve the equations of
motion coming from the action
\beq
S_{IIB}=\frac{1}{G_{10}}\int d^{10}x \sqrt{-g}\Big[
R-\frac{1}{2}(\partial\phi)^2 -\frac{e^{\phi}}{12}F_3^2  \Big],
\eeq
namely,
\bea
&&R_{\m\n}=\frac{1}{2}\pa_\m\f\pa_\n\f+\frac{1}{12}e^\f\left(3F_{\m\r_1\r_2}F_{}^{\r_1\r_2}{}_\n-\frac{1}{4}F_3^2 g_{\m\n}\right),
\NO\\
&&\nabla_\m\left(e^\f F_{}^{\m\r_1\r_2}\right)=0,\quad \pa_{[\m} F_{\r_1\r_2\r_3]}=0,\NO\\
&&\square_g\f-\frac{1}{12}e^\f F_3^2=0,
\label{eqsiib}
\eea
where we defined
\beqs
\nabla_\m\,X&\equiv&\partial_{\mu}\left(\sqrt{-g}\,X\right)\,,\nonumber\\
\square_g X&\equiv&\frac{1}{\sqrt{-g}}\partial_{\mu}\left(\sqrt{-g} g^{\mu\nu}\partial_{\nu}\,X\right)\,.\nonumber
\eeqs
These are the Einstein, Maxwell, and Klein-Gordon equations
in the curved background, and the Bianchi identity.
From here one derives the equations satisfied by  the
functions $[\phi(r), h(r),g(r),
k(r), a(r),b(r)]$ appearing in the background.
Each solution to these equations is conjectured to capture
the non-perturbative dynamics of the  field theory discussed above, either in a
different vacuum or after some operator is inserted in the
Lagrangian, deforming the theory (also in a given vacuum).

Looking for configurations preserving some SUSY is in practice easier
than looking for generic solutions of the
(second order) equations of motion (\ref{eqsiib}) for the background of eq. (\ref{nonabmetric424}),
since supersymmetric solutions are obtained by solving
first order ``BPS'' equations. It is usually the case, and it has been shown
explicitly for the background (\ref{nonabmetric424}), that solutions of the first order BPS equations
automatically solve the equations of motion.

\subsection{A useful result: The BPS equations}

The first order equations have been carefully derived in
\cite{Casero:2006pt}-\cite{HoyosBadajoz:2008fw}.
Let us quote a result from \cite{HoyosBadajoz:2008fw}: if we ``change
basis'' and write the
functions of the background in terms of a set of functions $P(\rho),
Q(\rho),Y(\rho), \tau(\rho), \sigma(\rho)$ as
\beq
4 e^{2h}=\frac{P^2-Q^2}{P\cosh\tau -Q}, \;\; e^{2g}= P\cosh\tau -Q,\;\;
e^{2k}= 4 Y,\;\; a=\frac{P\sinh\tau}{P\cosh\tau -Q},\;\; N_c b= \sigma.
\label{functions}
\eeq
Using these new variables, one can manipulate the BPS equations to obtain a
single decoupled second order equation for $P(\rho)$, while all other functions are
simply obtained from $P(\rho)$ as follows:
\bea
& & Q(\rho)=(Q_o+ N_c)\cosh\tau + N_c (2\rho \cosh\tau -1),\nonumber\\
& & \sinh\tau(\rho)=\frac{1}{\sinh(2\rho-2\rho_o)},\quad \cosh\t(\r)=\coth(2\r-2\r_o),\nonumber\\
& & Y(\rho)=\frac{P'}{8},\nonumber\\
& & e^{4\phi}=\frac{e^{4\phi_o} \cosh(2\rho_o)^2}{(P^2-Q^2) Y
\sinh^2\tau},\nonumber\\
& & \sigma=\tanh\tau (Q+N_c)= \frac{(2N_c\rho + Q_o + N_c)}{\sinh(2\rho
-2\rho_o)}.
\label{BPSeqs}
\eea
The second order equation mentioned above reads,
\beq
P'' + P'\Big(\frac{P'+Q'}{P-Q} +\frac{P'-Q'}{P+Q} - 4 \coth(2\rho-2\rho_o)
\Big)=0.
\label{master}
\eeq
So, to summarize, any solution of the eq.(\ref{master}) will generate via
the expressions in eq.(\ref{BPSeqs}) a set of functions $[h,g,k, \phi, a,
b]$ as obtained using eq.(\ref{functions}), such that when plugged back
into
eq.(\ref{nonabmetric424}) gives a string background
solving eq.(\ref{eqsiib}), preserving four
supercharges, that is dual to a minimally SUSY 4-d field theory.
%%%%%%%%%%%%%%%%%%%%%%%%%%%%%%%%%%%%
%%%%%%%%%%%%%%%%%%%%%%%%%%%%%%%%%%%%
%\newpage
\section{The approximate solution }

In this section we present a solution to the above BPS equations which
has the property that a suitably defined gauge
coupling -see below- possesses a plateau
in some intermediate energy scale. Since the solution is not known analytically in closed form, we proceed by
first exhibiting an approximate solution with
this property and then, in the next section we will show
how one can systematically compute the corrections to
this solution in the form of an expansion
in the coupling of a dimension-six operator.

Let us start with a few comments on the constants,
$\r_o$ and $Q_o$, appearing in the
equations (\ref{BPSeqs}) of the previous section. The constant $\r_o$
sets
the minimum possible endpoint of the geometry in the IR.
Without loss of generality we will set $\r_o=0$ in the following. The value of the constant $Q_o$ determines whether
the solution in fact extends up to $\r_o$ in the IR. Unless $Q_o=-N_c$, all
solutions will end in the
IR at some $\r_{IR}>\r_o$  \cite{HoyosBadajoz:2008fw}. Here we will consider only solutions with $Q_o=-N_c$ such that the solution
extends all the way to $\r_o=0$ in the IR.

In order to exhibit the approximate solution we are interested in, we write the
second order equation (\ref{master}) in the form
 \be\label{master-c+}
\pa_\r\left(s(P^2-Q^2)P'\right)+4sP'QQ'=0,
\ee
where
\be
s(\r)=\sinh^2\t=\frac{1}{\sinh^2(2\r)}.
\ee
Integrating eq.~(\ref{master-c+}) twice we obtain
\be\label{master-c+-integrated}
P^3-3Q^2P+6\int_{\r_2}^\r d\r'QQ'P+12\int_{\r_2}^\r d\r' s^{-1}\int^{\r'}_{\r_1} d\r'' s P'QQ'=c^3 R(\r)^3,
\ee
where
\be
R(\r)\equiv\left(\cos^3\a+\sin^3\a(\sinh(4\r)-4\r)\right)^{1/3},
\ee
and $c$, $\a$ are arbitrary integration constants, and we will fix the values of $0\leq \r_1, \r_2 \leq \infty$ later.
The approximate solution we are looking for is obtained by taking $c$ large compared to $N_c$. In this limit one
obtains the approximate solution
\be\label{approx_sol}
P\approx c R(\r).
\ee
An interesting property of this solution is that it makes the dilaton constant (to leading order in $N_c/c$),
namely
\be
e^{4(\f-\f_o)}=\frac{3}{c^3\sin^3\a}+\co\left((N_c/c)^2\right).
\ee

However, for this to be a well defined solution we need to ensure that $P(\r)> Q(\r)$ for all $\r\geq 0$,
since any $\r$ where $P=Q$ is a singular point of eq.~(\ref{master}) or eq.~(\ref{master-c+}).
Now, for small $\r$ we have
$Q(\r)=\co(\r^2)$, while $P(\r)\approx c\cos\a+\co(\r^3)$ and so $P>Q$ is ensured by requiring $\cos\a>0$.
For large $\r$ we have $Q(\r)\sim 2 N_c \r$, while $P(\r)\sim 2^{-1/3} c \sin\a e^{4\r/3}$.
So, requiring that $\sin\a >0$ is again sufficient to ensure that $P>Q$ for large $\r$. To ensure that
$P>Q$ for all $\r$, though, we need to look closer at the solution. Clearly, in the approximation of eq.~(\ref{approx_sol})
$P\geq c\cos\a$ for all $\r$. At some value of $\r=\r_*$, we have $\sinh(4\r_*)-4\r_*\approx \cot^3\a$ and $P$ starts deviating
from the constant value $c\cos\a$. As we shall see below, to allow for a large region of walking behavior we will need to take
$\cot\a\gg 1$, in which case
\be
\r_*\approx \frac14\left(\log 2+3\log\cot\a\right)\,\gg\,1.
\ee
In order to ensure that $P>Q$ everywhere, it is sufficient to require that $P(\r_*)> Q(\r_*)$, which
puts an upper bound on $\cot\a$, namely
\be\label{approx}
1\,\ll\,\cot\a \lesssim \exp\left(\frac{2^{4/3}}{3}\frac{c}{N_c}\right).
\ee
This relation defines the approximation we will be working with in the rest of the paper. In this approximation
$P(\r)$ remains almost constant for $0\leq \r\lesssim \r_*$.
As we will see below, the fact that $P$ is almost constant up to very large
scales $\r_*$ produces an intermediate energy
region over which the four-dimensional gauge coupling is
almost constant.
The larger $\cot\a$, the wider this region  is, with the only limitation provided by
the upper bound on the value of $\cot\a$, which depends on the ratio $c/N_c$.

%%%%%%%%%%%%%%%%%%%%%%%%%%%%%%%%%%%%
%%%%%%%%%%%%%%%%%%%%%%%%%%%%%%%%%%%%
%\newpage
\section{A systematic expansion}

In the limit $c/N_c\to \infty$ the solution presented in the previous section is an exact solution
of the BPS equations. However, it is instructive to determine the subleading corrections to this solution
in an expansion in powers of $N_c/c$. Such a systematic expansion
was  constructed in \cite{HoyosBadajoz:2008fw},
where the constant $c$ was identified with
the coupling of a dimension-6 operator. The leading term in
the expansion for large $c$ presented in
Appendix B of \cite{HoyosBadajoz:2008fw} is the solution in eq.~(\ref{approx_sol})
presented in the previous section with $\cos\a=0$. Generalizing this expansion to the case $\cos\a\neq 0$
is straightforward, as one merely needs to replace eq.~(\ref{approx_sol}) as the leading solution in the
recursion relations that determine the subleading corrections.

Following \cite{HoyosBadajoz:2008fw} we write $P$ in a formal expansion in inverse powers of $c$ as
\be\label{solution}
P=\sum_{n=0}^\infty c^{1-n} P_{1-n}.
\ee
Inserting this expansion in eq.~(\ref{master-c+-integrated}) we obtain recursively
\bea
&&P_1=R,\NO\\%\NO\\
&&P_0=0,\NO\\%\NO\\
&&P_{-1}=-\frac13 P_1^{-2}\left(-3Q^2P_1+6\int_{\r_2}^\r d\r' QQ' P_1+12\int_{\r_2}^\r d\r' s^{-1}\int^{\r'}_{\r_1} d\r'' sQQ'P_1'\right),\NO\\\NO\\
&&P_{-2}=0,\NO\\%\NO\\
&&P_{-n-2}=-\frac13P_1^{-2}\left\{\sum_{m=1}^{n+2}\left(2P_1P_{1-m}P_{m-n-2}+\sum_{k=1}^{n-m+3}P_{1-m}P_{1-k}
P_{m+k-n-2}\right)-3Q^2P_{-n}\right.\NO\\
&&\left.+6\int_{\r_2}^\r d\r' QQ' P_{-n}+12\int_{\r_2}^\r d\r' s^{-1}\int^{\r'}_{\r_1} d\r'' sQQ'P_{-n}'\right\},\quad n\geq 1.
\eea
It follows by induction that $P_k=0$ for all even $k$.

At this point we have to make a choice for the values of $\r_1$ and $\r_2$. Given that $P_1\sim e^{4 \r/3}$ as $\r\to \infty$,
requiring that $P_k$ for $k<1$ are all subleading with respect to $P_1$ as $\r\to\infty$ sets $\r_1=\infty$.
 Moreover, provided $\cos\a\neq 0$, as $\r\to 0$,
\bea
Q&=&2N_c\left(\frac23 \r^2+\co(\r^4)\right),\NO\\
P_1&=&\cos\a\left(1+\frac{32}{9}\tan^3\a\r^3+\co(\r^5)\right).
\eea
Requiring then that $P_k$ for $k<1$ are all subleading with respect to $P_1$ as $\r\to 0$ sets $\r_2=0$. With these choices,
for odd $k$, $P_k\sim e^{4k\r/3}\r^{m(k)}$ as $\r\to \infty$, where $m(k)$ is a $k$-dependent positive integer, while
as $\r\to 0$, $P_k=\co(\r^3)$ for $k<1$. In particular, provided $\cos\a\neq 0$, for small $\r$
\be
P=c\cos\a+\m(c,\a) \r^3+\co(\r^4),
\ee
which shows that, provided $\cos\a\neq 0$, the solution described by eq.~(\ref{solution}) has type I IR asymptotics
(see eq.~(4.24) in \cite{HoyosBadajoz:2008fw}).
The corresponding  behavior in the IR (for $\r\rightarrow 0$)
of the functions in eq.~(\ref{functions}) is
\bea
&&e^{2h}=\frac12 c \cos\a \r\left(1-\frac{4\r^2}{3}+\frac{\m+8N_c/3}{c\cos\a}\r^3+\co(\r^4)\right),\NO\\
&&e^{2g}=\frac{c \cos\a}{2\r}\left(1+\frac{4\r^2}{3}+\frac{\m-8N_c/3}{c\cos\a}\r^3+\co(\r^4)\right),\NO\\
&&e^{2k}=\frac32\m\r^2+\co(\r^3),\NO\\
&&e^{4(\f-\f_o)}=\frac{32}{3\m c^2\cos^2\a}+\co(\r),\NO\\
&&a=1-2\r^2+\frac{8N_c}{3c\cos\a}\r^3+\co(\r^4),\NO\\
&&b=1-\frac23\r^2+\co(\r^4).
\eea

\begin{figure}[htp]
\begin{center}
\includegraphics{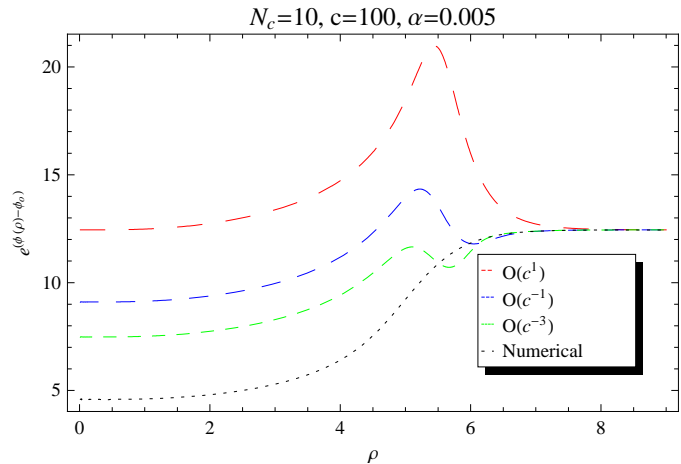}
\caption{Plot of the dilaton $\f$ as a function of $\r$. The first three orders in the expansion (\ref{solution})
are compared with the numerical solution.}
\label{Fig:dilaton}
\end{center}
\end{figure}

\begin{figure}[htp]
\centering
\scalebox{.87}{\includegraphics{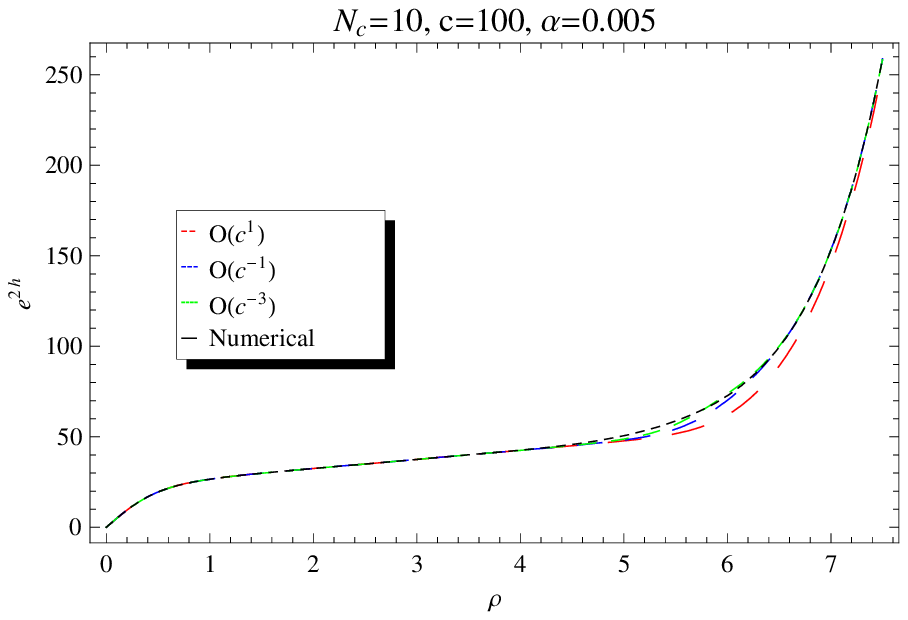}}
\hfill
\scalebox{.87}{\includegraphics{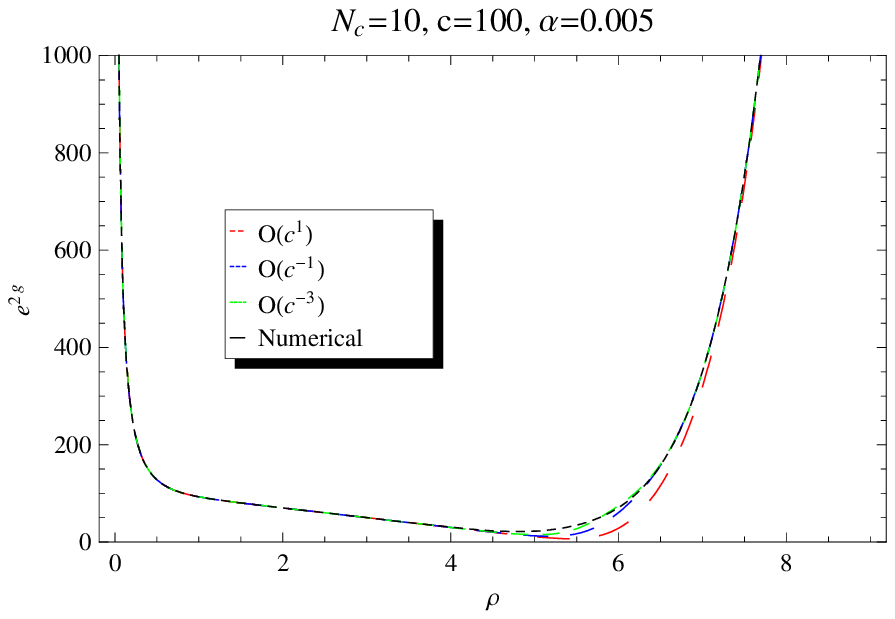}}\\
\scalebox{.87}{\includegraphics{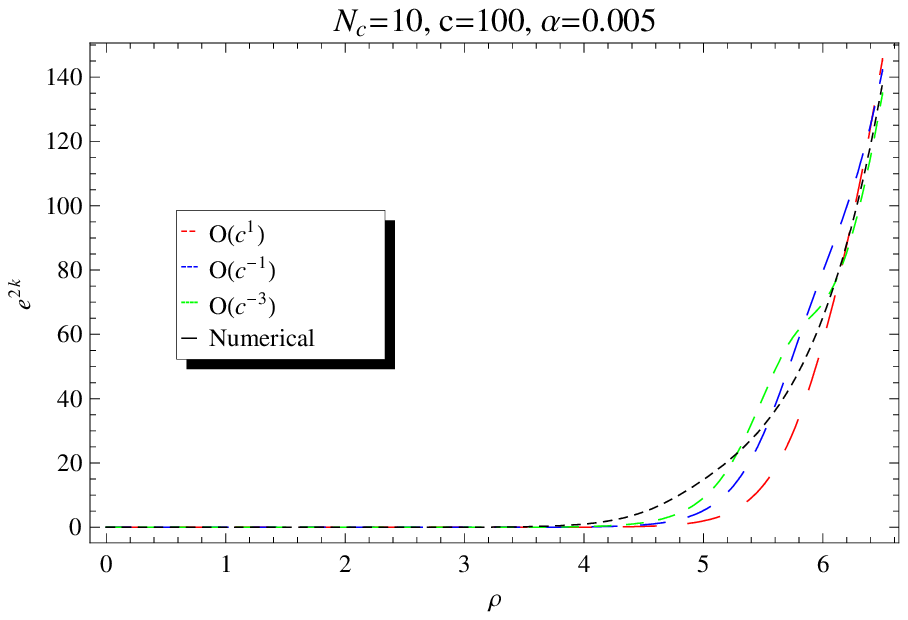}}
\hfill
\scalebox{.87}{\includegraphics{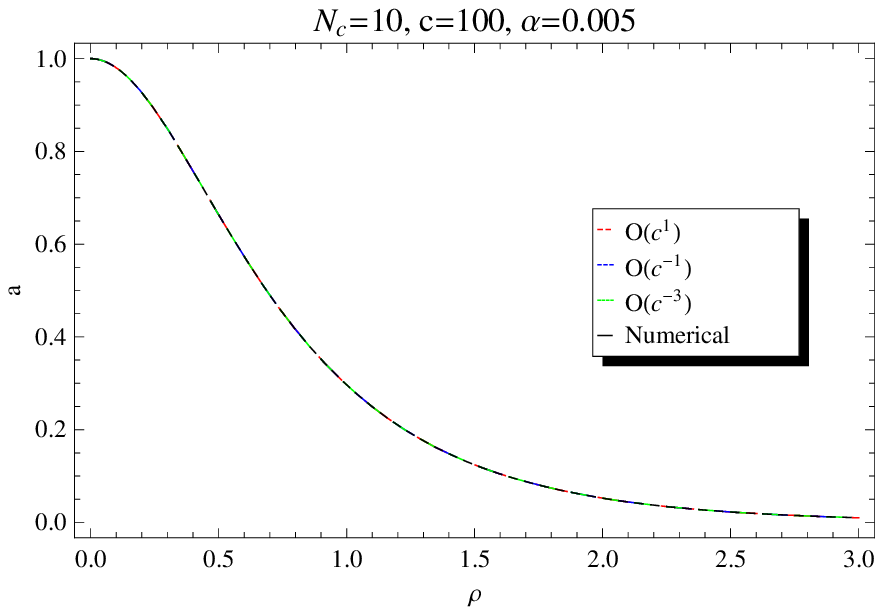}}
\caption{The background functions $h$, $g$, $k$ and $a$ as a function of $\r$. Here we plot the first three orders
in the expansion (\ref{solution}) and we compare it with the numerical solution. It is clear that the expansion
(\ref{solution}) converges sufficiently fast.}
\label{Fig:background}
\centering
\end{figure}

The constant $\m$ is a very non-trivial function of $c$ and $\a$ given by
\be
\m(c,\a)=\frac{16}{3\cos^2\a}\left(\frac{2}{3}c \sin^3\a+\sum_{k=0}^\infty c^{-2k-1}\int_0^\infty d\r's QQ'P'_{-2k+1}\right).
\ee
In the approximation eq.~(\ref{approx}) the first correction can be evaluated approximately to obtain
\be
\m(c,\a)\approx\frac{16}{3\cos^2\a}\left(\frac{2}{3}c \sin^3\a+\frac{N_c^2\sin^3\a}{3 c \cos^2\a}\left(\log 2+3\log\cot\a\right)^2
+\co(N_c^4/c^3)\right).
\ee
It is easy to see from eq.~(\ref{master-c+}) and the BPS equations eq.~(\ref{BPSeqs}) that $\m(c,\a)$ is related to the
value of the dilaton at $\r=0$ (see also Fig.~\ref{Fig:dilaton}), namely
\bea
e^{4(\f(0)-\f_o)}&=&\frac{32}{3\m(c,\a)c^2\cos^2\a}\NO\\
&\approx&\frac{3}{c^3\sin^3\a}
\left(1-\frac12\left(\frac{\log 2+3\log\cot\a}{\cos\a}\right)^2\left(\frac{N_c}{c}\right)^2+\co\left((N_c/c)^4\right)\right).
\eea
Numerical calculations show that this is indeed a very good approximation.
For the UV (large $\rho$) expansions, we refer the reader to the Class II
asymptotics in Section 4.3.1 of the paper \cite{HoyosBadajoz:2008fw}. Plots of the
background functions as functions of $\r$ are shown in Figs.~\ref{Fig:dilaton},~\ref{Fig:background}.

%%%%%%%%%%%%%%%%%%%%%%%%%%%%%%%%%%%%
%%%%%%%%%%%%%%%%%%%%%%%%%%%%%%%%%%%%
%\newpage
\subsection{A short discussion on the singularity}
The aim of this short section is to briefly summarize the properties of the
singularity that our IIB solution presents in the IR (that is, for small
values of the radial coordinate).
This paper is mostly concerned with the properties of the metric away from the
singularity (for $\rho> 1$), where  walking dynamics seems to emerge,
yet it is worth reminding the reader about the characterization of
the singularity as {\it good}, and hence not problematic.

The Ricci
scalar (and other invariants) diverge at $\rho=0$.
Accepting such a singularity in the background
requires to specify in what sense and
 to what extent the AdS/CFT ideas apply.
 This problem did not go unnoticed and since the early days different
criteria were developed, in order to select when a singular space-time
can be accepted or must be rejected as the dual
to a field theory.

The outcome of these studies
is a set of criteria, which if satisfied ensure  that the strongly-coupled dynamics of
the field theory is faithfully reproduced by the string solution.
We call  {\it good} singularities those that satisfy these criteria.
One of these criteria requires that the potential of the
supergravity theory (after reduction to five-dimensions) be bounded from above
\cite{Gubser:2000nd}.
A second criterion requires that the component $g_{tt}$  of the metric be
bounded~\cite{Maldacena:2000mw}.
In many examples, these two criteria give the same
result, accepting or rejecting the same singular space-times, though
no general proof is known about the equivalence of these two criteria.
In our case,  $|g_{tt}|=e^{\phi}$  is
bounded and hence our IR singularity is {\it good},
satisfying the criterion in~\cite{Maldacena:2000mw}.

One characteristic of  {\it good} singularities is the fact
that the calculation of many invariants involving the singular region
yields finite results.
For instance, in  computing the
invariant $\sqrt{- g}$ the singularities at $\rho\rightarrow 0$
cancel .
Other examples of this cancellation involve certain
brane probe actions, and this fact is believed to be
robust enough that the presence of a {\it good} singularity
should not be a reason of concern.

%%%%%%%%%%%%%%%%%%%%%%%%%%%%%%%%%%%%
%%%%%%%%%%%%%%%%%%%%%%%%%%%%%%%%%%%%
%\newpage
\section{About the gauge coupling}
\label{gauge-coupling}
%{\it Here we discuss how to define the gauge coupling of our 4-d QFT. This
%section is targeted mostly to people not working on String Theory.}

%A natural question is
%how to define different physical quantities.
The next step is to identify the 4-d physical quantities and to relate them to the background.
For example, defining the gauge coupling and its running with energy. In this
section we comment on one aspect of this problem.

It will be convenient in this section to reinstate the factors of
$\alpha'$ and $g_s$.
Let us start by reminding the reader that the six-dimensional theory on
the D5 branes
has a 't Hooft coupling given by $\lambda_6=g_s \alpha' N_c$. This
coupling has units of length squared as a 6-d coupling should. Now, when
we
wrap the  branes on a small two-cycle $\Sigma_2$ and explore the theory at
low
energies, we will effectively observe a 4-d theory. A natural way of
defining the (dimensionless) gauge coupling of this theory is by combining,
\beq
g_{4}^2= \frac{g_{6}^2}{Vol \Sigma_2}.
\eeq
Then, the question is how to select the cycle $\Sigma_2$.
It can be shown that the two-manifold defined by the surface
\beq
\Sigma_2=[\theta=\tilde{\theta},\;\;
\varphi=2\pi-\tilde{\varphi},\;\;\psi=\pi],
\label{2cycle}
\eeq
with the other coordinates $x^\mu,\rho$ taken to be constant, defines the
only two-cycle in
the geometry of eq.(\ref{nonabmetric424})\footnote{Actually, there is another possible two
cycle, given by $\Sigma_2'=[\theta=\pi-\tilde{\theta},\;
\varphi=\tilde{\varphi}\;\; \psi=\pi]$, but this one will give the same
results for all the observables we consider here. Probably it is the same
two-cycle, with a different orientation.}.
 The characteristic of the submanifold
in eq.(\ref{2cycle}) is that it gets rid of the `fibration terms'.
In turn, this means that this is the only dimension-2 manifold (without
boundaries) that the
branes can wrap.

A good definition of the 4-d coupling can be obtained following the
papers \cite{DiVecchia:2002ks} that considered a ``probe'' five brane
extended along the Minkowski directions and the two-cycle of
eq.(\ref{2cycle}) $R^{1,3}\times \Sigma_2$ in
the background eq.~(\ref{nonabmetric424}).

The authors of \cite{DiVecchia:2002ks} turn on a gauge field $F_{\mu\nu}$
on
the Minkowski directions of this five-brane probe (with tension $T_{D5}$)
and compute its action
(given by
the Born-Infeld-Wess-Zumino action);
\beq
S_{probe}= -T_{D5} \int d^4 x d\Sigma_2 e^{-\phi}\sqrt{-\det[g_{\mu\nu,ind}
+2\pi
\alpha'
F_{\mu\nu}]} + T_{D5}\int C_{2,ind}\wedge F_2\wedge F_2
\label{biwz}
\eeq
The induced 6-d configuration on the probe brane extends along
$R^{1,3}\times \Sigma_2$ is
\bea
& & ds_{ind}^2= e^{\phi}\Big[dx_{1,3}^2 +\alpha' g_s \Big(
e^{2h}+\frac{e^{2g}}{4}(a-1)^2    \Big)(d\theta^2+\sin^2\theta d\varphi^2)
\Big],\nonumber\\
& & C_{2,ind}= \frac{N_c}{2}(\psi-\psi_0)\sin\theta d\theta \wedge
d\varphi,\nonumber\\
& & F_2= F_{\mu\nu} dx^\mu \wedge dx^\nu .
\label{d5induced}
\eea
From here we can compute the determinant\footnote{Notice that the
term $\int
C_{2,ind}\wedge F_2\wedge F_2$ will generate the theta-term $\Theta \int
d^4 x F_{\mu\nu} {}^{*}F^{\mu\nu}$. We will omit this term in the following
and concentrate on the definition of the gauge coupling.}
\beq
\sqrt{\Big( -\det[g_{ab,ind}+2\pi \alpha' F_{ab}]\Big)}=e^{3\phi}
(\alpha'g_s ) [e^{2h}+\frac{e^{2g}}{4}(a-1)^2]\sqrt{1 - 4\pi^2\alpha'^2
F_{\mu\nu}^2 }\sin\theta .
\eeq
Plugging this result into eq.(\ref{biwz}) and expanding for small values
of the field strength (or equivalently, for $\alpha'\to 0$), we get
\bea
& & S_{probe}\approx -T_{D5} \int d^4 x \int_{0}^{\pi}\sin\theta d\theta
\int_{0}^{2\pi}d\varphi \times \nonumber\\
& & \Big[
e^{2\phi}(\alpha'g_s ) [e^{2h}+\frac{e^{2g}}{4}(a-1)^2]\big(1- 4\pi^2
\alpha'^2 e^{-2\phi}\eta^{\mu\alpha}\eta^{\nu\beta}F_{\mu\nu}F_{\alpha\beta}
\big)\Big],
\eea
where we used that  $g^{\mu\nu}= e^{-\phi}\eta^{\mu\nu}$. Now, performing
the
integral over the
angles, the term with the Yang-Mills
action reads,
\beq
S_{probe}\approx 4\pi
T_{D5}(2\pi^2\alpha'^2)(\alpha' g_s )[e^{2h}+\frac{e^{2g}}{4}(a-1)^2]\int
d^4 x F_{\mu\nu} F^{\mu\nu}=\frac{1}{4 g^2}\int
d^4 x F_{\mu\nu} F^{\mu\nu}.
\eeq
From here, and using the fact that the tension of a five brane satisfies $32 \pi^5
g_s
\alpha'^3 T_{D5}=1$, we read that the 4-d gauge coupling
is
\beq
\frac{8\pi^2}{g^2 }= 2 [e^{2h}+\frac{e^{2g}}{4}(a-1)^2]= P e^{-\tau}.
\label{coupling}
\eeq
Before we evaluate this quantity on our solution, a few comments are in
order.

i) It was shown in the paper \cite{Nunez:2003cf},
that the five branes on the
submanifold
$R^{1,3}\times \Sigma_2$ preserve SUSY only if the probe brane is at
infinite radial distance from the end of the space ($\rho\to \infty$).
This result is valid for the particular solution considered there and does
not extend to the present background. Indeed, a five brane in the
configuration described above does not preserve the same spinors as the
background, hence, it is not SUSY, rendering our definition of gauge
coupling non SUSY preserving. This will not concern us and we will take
the definition of
eq.(\ref{coupling}) as a good estimation, valid over all the range of the
radial coordinate.

ii) If we consider this definition, we can see that the gauge coupling
diverges in the IR (signaling confinement). Also, this gauge coupling
vanishes in the far UV. The fact that it becomes small should not be
a reason of worry. This 4-d gauge being small is not indicating that the ten
dimensional geometry is highly curved. Indeed, the 4-d coupling becomes
small when the geometry is smooth and well approximated by IIB supergravity.

iii) Using a particular radius-energy relation it was shown in
\cite{DiVecchia:2002ks} that one particular solution to the eq. of motion
(\ref{master}), reproduces the NSVZ beta function. For the class of
solutions we are interested in, the paper \cite{HoyosBadajoz:2008fw}
shows that the beta function is characteristic of a 6-d field theory. %The
%difference between both behaviors was explained in
%\cite{HoyosBadajoz:2008fw}.

In Figure 3 we plot the 4-d gauge coupling as defined in
eq.(\ref{coupling}) on the solutions we found above.

%HERE WE PUT A PLOT OF $g^2$

\begin{figure}[htp]
\begin{center}
\scalebox{.75}{\includegraphics{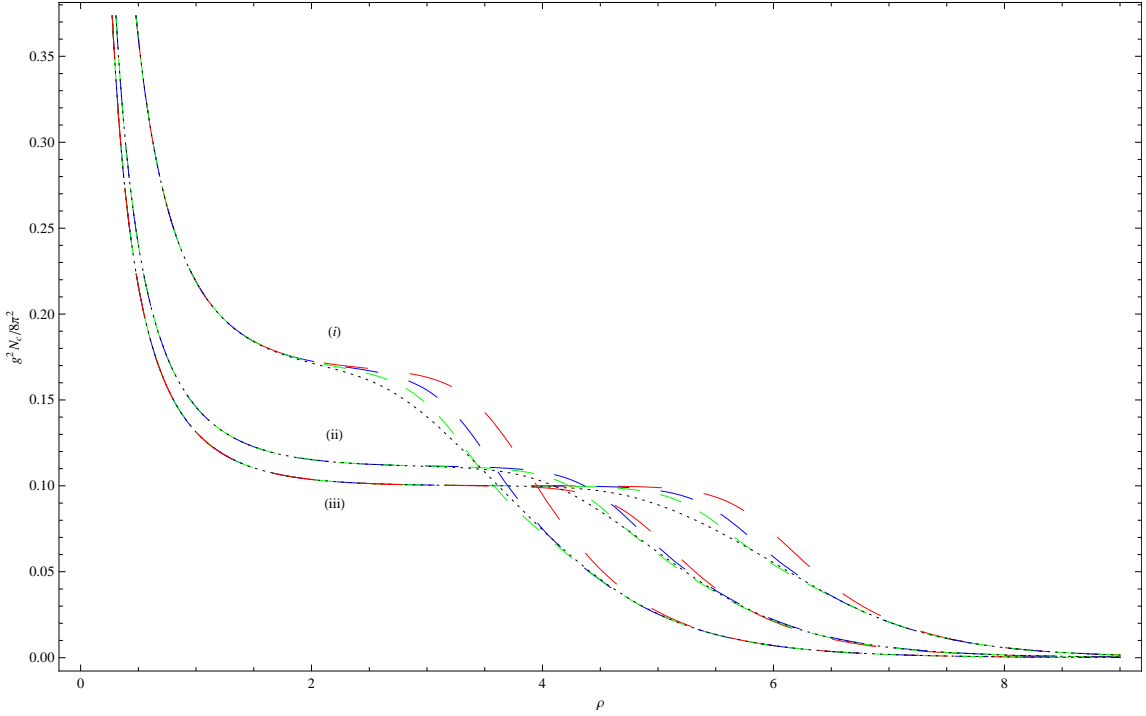}}
\caption{The 't Hooft coupling $g^2N_c/(8\pi^2)$
as a function of $\rho$ for various values of the parameters $c$, $\a$. All three curves are for $N_c=10$, while
$c=60$, $\a=0.01$ for (i), $c=90$, $\a=0.002$ for (ii) and $c=100$, $\a=0.0005$ for (iii). The red (long dashes) curves
are the $\co(c)$ approximation in the expansion (\ref{solution}), the blue (medium dashes) lines are the $\co(1/c)$ approximation,
the green (short dashes) lines are the $\co(1/c^3)$ approximation, and the black (dotted) lines are the numerical solutions.}
\label{Fig:plotcoupling}
\end{center}
\end{figure}

%\begin{figure}[ht]
%\begin{center}
%\includegraphics[width=0.8\linewidth]{plotcouplingx0.eps}
%\caption{The 't Hooft coupling $g^2N_c/(8\pi^2)\times 4\rho_{UV}$
%as a function of $\rho$, obtained by choosing of $c_2=2\bar{c_2}$.
%The red curves are the numerical solutions, in black the approximate
%solutions.
%The three curves correspond to choices of $c_1$
%such as to have $\rho_{UV}=12,7,5,2$ (right to left).}
%\label{Fig:plotcouplingx0}
%\end{center}
%\end{figure}

%\newpage
\section{Approximate symmetries of the solution}

As we have seen above, the gauge coupling exhibits three qualitatively different regimes.
Namely, the IR regime for $0<\r \lesssim 1$, an intermediate plateau for $1 \lesssim \r \lesssim \r_*$
where the gauge coupling is almost constant, and a UV regime for $\r \gtrsim \r_*$. We study the (approximate) symmetries of the background in these three different
regimes.

To leading order in $c/N_c$, i.e. in the approximation of eq.~(\ref{approx_sol}), the background of eq.(\ref{nonabmetric424})-using the
definitions in eq.(\ref{functions})-takes the form
\bea
ds^2&=&\frac{\sqrt{3}}{c^{3/2}\sin^{3/2}\a}\left\{dx_{1,3}^2+\frac{cP_1'}{8}\left(4d\r^2+(\om_3+\tilde{\om}_3)^2\right)\right.\NO\\
&&\left.+\frac{cP_1\cosh\t}{4}\left(d\Om_2^2+d\tilde{\Om}_2^2+2\tanh\t\left(\om_1\tilde{\om}_1-\om_2\tilde{\om}_2\right)\right)
+\co\left(N_c^2/c\right)\right\},
\eea
\be
F\sub{3}=N_c\left\{-d\left[\frac{2\r}{\sinh(2\r)}(\om_1\wedge \tilde{\om}_1-\om_2\wedge \tilde{\om}_2)\right]
+\frac14\left(\om_1\wedge\om_2-\tilde{\om}_1\wedge\tilde{\om}_2\right)\wedge(\om_3+\tilde{\om}_3)\right\},
\ee
where $d\Om_2^2=\om_1^2+\om_2^2=d\th^2+\sin^2\th d\vf^2$ and
$d\tilde{\Om}_2^2=\tilde{\om}_1^2+\tilde{\om}_2^2=d\tilde{\th}^2+\sin^2\tilde{\th}d\tilde{\vf}^2$ are the metrics
on two 2-spheres, and we have introduced the one-forms
\be
\om_1=d\th,\quad \om_2=\sin\th d\vf,\quad \om_3=\cos\th d\vf,
\ee
in addition to the left-invariant $SU(2)$ forms of eq.~(\ref{su2}). Let us now look at the behavior of this background in the
three different dynamical regimes.

Starting with the far IR, $\r \lesssim 1$, we see from Table \ref{regimes} that no further simplifications
occur in the above background and so the only isometries are the obvious $ISO(1,3)\times U(1)\times U(1)\times \mathbb{Z}_2$, being the two $U(1)'s$
the traslations in $\varphi, \tilde{\varphi}$, while $\mathbb{Z}_2$ comes from the change $\psi\to \psi+2\pi$.

In the intermediate regime, $1 \lesssim \r \lesssim \r_*$, we see that the background simplifies
to
\be
ds^2\approx \frac{\sqrt{3}}{c^{3/2}\sin^{3/2}\a}\left\{dx_{1,3}^2
+\frac{c\cos\a}{4}\left[\left(\frac{\tan^3\a e^{4\r}}{3}\right)\left(4d\r^2+(\om_3+\tilde{\om}_3)^2\right)
+d\Om_2^2+d\tilde{\Om}_2^2\right]\right\},
\label{Eq:intermediate}
\ee
\be
F\sub{3}\approx\frac{N_c}{4}\left(\om_1\wedge\om_2-\tilde{\om}_1\wedge\tilde{\om}_2\right)\wedge(\om_3+\tilde{\om}_3).
\ee
Note that since $\r< \r_*$ in this regime we have $\tan^3\a e^{4\r}/2 <1$. If this was equal to $1$ then the
compact part of the metric would be that of $T_{11}$, the base of the conifold. This indeed happens in the UV region,
$\r \gtrsim \r_*$, where the background takes the form
\be
ds^2\approx \frac{\sqrt{3}}{c^{3/2}\sin^{3/2}\a}\left\{dx_{1,3}^2
+\frac{2^{-1/3}c\sin\a}{4}e^{4\r/3}\left(\frac23\left(4d\r^2+(\om_3+\tilde{\om}_3)^2\right)
+d\Om_2^2+d\tilde{\Om}_2^2\right)\right\},
\label{Eq:farUV}
\ee
\be
F\sub{3}\approx\frac{N_c}{4}\left(\om_1\wedge\om_2-\tilde{\om}_1\wedge\tilde{\om}_2\right)\wedge(\om_3+\tilde{\om}_3).
\ee
It can be shown that the metric
\be
ds^2=d\Om_2^2+d\tilde{\Om}_2^2+\z(\om_3+\tilde{\om}_3)^2,
\ee
where $\z$ is a constant,
and the 3-form given above possess an $SU(2)\times SU(2)\times U(1)$ isometry {\em for any non-zero value of $\z$}. In
particular, irrespective of the value of $\z$, the seven Killing vectors take the form
\[
 \begin{array}{ll}
\xi_1=\sin\vf \pa_\th+\cot\th \cos\vf\pa_\vf-\csc\th\cos\vf\pa_\psi &
\tilde{\x}_1=\sin\tilde{\vf} \pa_{\tilde{\th}}+\cot\tilde{\th} \cos\tilde{\vf}\pa_{\tilde{\vf}}
-\csc\tilde{\th}\cos\tilde{\vf}\pa_\psi \\
\xi_2=-\cos\vf \pa_\th+\cot\th \sin\vf\pa_\vf-\csc\th\sin\vf\pa_\psi &
\tilde{\x}_2=-\cos\tilde{\vf} \pa_{\tilde{\th}}+\cot\tilde{\th} \sin\tilde{\vf}\pa_{\tilde{\vf}}
-\csc\tilde{\th}\sin\tilde{\vf}\pa_\psi \\
\xi_3=\pa_\vf & \tilde{\x}_3=\pa_{\tilde{\vf}}\\
\multicolumn{2}{c}{\xi=\pa_\psi}
\end{array}
\]
It follows that both in the intermediate regime, $1 \lesssim \r \lesssim \r_*$, and in the far UV regime, $\r \gtrsim \r_*$,
the background possess an approximate $ISO(1,3)\times SU(2)\times SU(2)\times \mathbb{Z}_{2N_c}$ isometry, where
the $U(1)$ generated by $\pa_\psi$ is broken to $\mathbb{Z}_{2N_c}$ by the R-symmetry anomaly (see for example \cite{Gursoy:2003hf}).

The Killing vectors are the same for the metrics
in the intermediate region ($1<\rho<\rho_*$) given in eq.~(\ref{Eq:intermediate}) and
that in the ``UV-region'' ($\rho>\rho_*$) written
in eq.~(\ref{Eq:farUV}). Nevertheless, there are differences between
the metrics in eq.~(\ref{Eq:intermediate}) and in eq.~(\ref{Eq:farUV}).
Indeed, ignoring the Minkowski part,
the metric in eq.~(\ref{Eq:intermediate}) can be intuitively pictured as
a cigar-shaped geometry, with a radial direction $\rho$,
a `circle' $(\omega_3 +\tilde{\omega}_3)$ all fibered over the $S^2 \times S^2$ manifold.
On the other hand, the metric of eq.~(\ref{Eq:farUV}) is the conifold, that can be intuitively
thought of as a cone over $S^2 \times S^3$. This in turn is responsible for an important distinction.
In the metric of eq.~(\ref{Eq:farUV}) we can rescale
the coordinates
\begin{equation}
x_i \to \delta x_i, \;\;\;\; \rho \to \rho+\frac{3}{2}\log\delta
\end{equation}
and the metric of eq.~(\ref{Eq:farUV}) is just conformal to itself
(that is it gets multiplied by a factor of $\delta^2$). This may suggest
that the theory gains some kind of scale-invariance
when approaching the far UV
region.
This is reminiscent of the picture proposed in \cite{Kazakov:2002jd}.
Notice also that due to subleading corrections in $N_c/c$ 
the dilaton changes at $\r_{\ast}$, as can be seen in Fig.~\ref{Fig:dilaton}.

To close this section, let us mention that the fact that the metric and
the background in the three regions
have a different geometric interpretation, and different symmetries,
implies that there exist three distinct dynamical phases in the dual
field theory,
and that the scales separating them have a physical meaning
that goes beyond the analysis of the gauge coupling we focused upon
in this paper.

In order to better understand what the intermediate phase is,
it would be interesting to study the model in limit in which
the gaugino condensate is switched off, to learn about the phase structure
of the underlying
six-dimensional gauge theory, also in relation to the
compactification
of the internal space. Finally to study the spectrum of the
theory and how the dynamical scales determine it.
All of this is going to be explored elsewhere.
\begin{table}
\[
\begin{array}{|c|c|c|c|}
\hline\hline &&&\\
\r &  0<\r \lesssim 1 & 1 \lesssim \r \lesssim \r_* & \r \gtrsim \r_* \\&&& \\ \hline \hline&&&\\
P & c\cos\a & c\cos\a & c 2^{-1/3}\sin\a e^{4\r/3} \\&&&\\ \hline &&&\\
Q & N_c(2\r \coth(2\r)-1) & 2N_c\r & 2N_c\r  \\&&&\\ \hline &&&\\
e^{4(\f-\f_o)} & \frac{3}{c^3\sin^3\a}\left(1+\co((N_c/c)^2)\right) & \frac{3}{c^3\sin^3\a}\left(1+\co((N_c/c)^2)\right) & \frac{3}{c^3\sin^3\a}  \\&&&\\ \hline &&&\\
Y & \frac13 c\cos\a\tan^3\a\sinh^2(2\r) & \frac{1}{12}c\cos\a\tan^3\a e^{4\r} & \frac16 c 2^{-1/3}\sin\a e^{4\r/3} \\&&& \\ \hline &&&\\
\sinh\t & \frac{1}{\sinh(2\r)} & 2 e^{-2\r} & 2 e^{-2\r} \\&&& \\ \hline &&&\\
\cosh\t & \coth(2\r) & 1 & 1 \\&&& \\ \hline\hline
\end{array}\]
\caption{The leading behavior of the background functions in the three different dynamical regimes.}
\label{regimes}
\end{table}

%%%%%%%%%%%%%%%%%%%%%%%%%%%%%%%%%%%
%%%%%%%%%%%%%%%%%%%%%%%%%%%%%%%%%%%
\newpage
\section{Operator analysis}

In order to gain some insight on the meaning of this class of solutions
for the dual field theory, it is useful to examine the behavior of the leading-order approximation
$P\simeq P_1 = c R$ in the far UV, for $\r \gg \r_*$.
We can change variables, $\rho=-\frac{3}{2}\log z$,
where $z$ is now proportional to a length scale.
The physical scales in the problem can be rewritten in terms of $0<z<1$.
The far IR scale at which the space ends $\r=\r_0=0$ corresponds to $z_0=1$.
The scale $\r\sim 1$  corresponds to $z_1=\exp[-2/3]\simeq 1/2$.
The scale $\r_*\gg 1$ corresponds to $z_*=\exp[-2/3\r_*] \ll 1$,
and finally the far UV in which the background approaches the conifold
is $z\rightarrow 0$.

Expanding around $z\rightarrow 0$:
\beqs
P&\simeq&\frac{c\sin{\alpha}}{2^{1/3}}\frac{1}{z^2}\,+\,
\frac{2^{2/3}c \cos^3{\alpha}}{3\sin^2{\alpha}}z^4\,+\,
{2^{5/3}c\sin{\alpha}}z^4 \log z\,+\,O\left(\frac{N_c}{c}\right)\,\\
&\simeq&
\frac{c\sin\alpha}{2^{1/3}}\left(\frac{1}{z^2}\,+\,
\frac{2\cot^3\alpha}{3}z^{4+{6}{\tan^3\alpha}}\right)\,,\\
a&\simeq&2z^3\,+\,O\left(\frac{N_c}{c}\right)\,.
\eeqs
The expressions for the other functions appearing in the background are similar.

In this expansion, one can identify the presence of three independent
dynamical quantities. The
 $z^3$ term signals the presence of a dimension-3 condensate
 (usually interpreted as gaugino condensate), responsible for the behavior in
 the IR at the scale $z_0=1$.
 The coefficient $2^{-1/3}c\sin\alpha $ appears
 in front of the $z^{-2}$ term, which can be thought of as
 the effect of deforming the theory with the insertion of an
 operator of dimension-6, with non-vanishing coupling.
The presence of the term that, up to logarithmic corrections,
scales as $z^4$ can be thought of as the VEV of an operator
that is dimension-4 in the limit in which $\cot^3\alpha\gg 1$.
The size of the VEV is given by $2\cot^3\alpha/3$.

The interpretation of the running of the gauge coupling is hence
that at very high energy the model is dominated by the
insertion of the dimension-6 operator.
At the scale $z_*$ (equivalently, $\r_*$) the dimension-4 condensate appears and dominates
the dynamics. This operator is marginally irrelevant, and hence its influence
is superseded at low energies (below $\r \lsim 1$, or at distance $z>z_1$)
by the dimension-3 operator.
The operators of dimension 3 and 6 dominate the dynamics at small and large energies,
respectively. The existence of a finite intermediate range governed by the dimension 4 operator
results from its coefficient being large.

Notice that we could approximate the expression for $P$ by reabsorbing the $\ln z$
correction in the scaling dimension of the VEV.
In doing so, using the expression for $\r_*$ one sees that the VEV scales as $z^d$, where
\beqs
d=4+12e^{-4\r_*}\,.
\eeqs
This is interesting for two reasons. First,
because it shows that not taking $\cot\alpha \gg 1$,
would correspond to giving a VEV to
a highly irrelevant operator. As a result, the background
would be effectively governed just by two effects: the
dimension-6 coupling and the dimension-3
VEV. This is the case already  studied in the literature~
\cite{HoyosBadajoz:2008fw},
and we will not comment any further on it.
The second interesting fact is that
in the intermediate energy region $1\lsim\r\lsim\r_*$ the running is governed by a quasi-marginal
operator with scaling dimension $d\simeq 4$: the length of the intermediate region is
governed by the difference $d-4$, as  reasonable to expect.

Reinstating the proper units leads to the replacement
\beqs
\rho\,%\rightarrow \, \rho-\rho_0
&=& \frac{3}{2}\log\frac{\mu-\Lambda_0+\Lambda}{\Lambda}\,,\\
\mu&=&\left(e^{\frac{2}{3}%(
\rho%-\rho_0)
}-1\right)\Lambda+\Lambda_0\,,
\eeqs
where $\Lambda$ and $\Lambda_0$ have
dimension of an energy, and where $\Lambda_0$ is the scale at which
the singularity in the solution appears. This is the scale of confinement.

\begin{figure}
\begin{center}
\psfrag{a}{\large{$\L_0$}}
\psfrag{b}{\large{$\L_{IR}$}}
\psfrag{c}{\large{$\L_*$}}
\psfrag{d}{\large{$\m$}}
\psfrag{1}{\large{$<\co_3>$}}
\psfrag{2}{\large{$<\co_4>$}}
\psfrag{3}{\large{$\co_6$}}
\scalebox{.75}{
\includegraphics{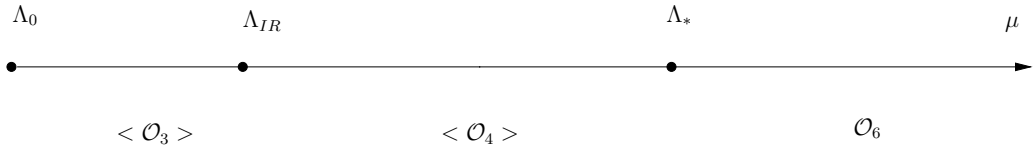}
}
\caption{The three qualitatively different energy regimes and the corresponding operators that dominate
the solution in each of these regions.}
\label{regimes-graph}
\end{center}
\end{figure}

The gaugino condensate is proportional to the scale $\Lambda^3$
and it dominates the dynamics for scales below $\r_{IR}\simeq 1$, that is
for $\mu <\L_0+\L$. The scale $\rho_{IR}\simeq 1$ at which the dimension-4 operator
takes over in governing the running corresponds to a physical scale
 $\mu=\Lambda_{IR}\sim \Lambda_0 + \Lambda$.
The $\r_*$ scale at which the theory becomes controlled by the
dimension-6 operator
corresponds to a third, independent, dynamical energy
scale
\beqs
\Lambda_{*}&=&\Lambda_0
+\left(\exp\left[\frac{2}{3}%(
\rho_{*}%-\rho_0)
\right]-1\right)\Lambda\,\gg\,\Lambda_{IR}\,.
%,\\
%&\simeq&\Lambda_0+\Lambda \exp\left[\frac{3}{2}\rho_{UV}\right]\, ,
\eeqs
above which the gauge theory is effectively six-dimensional.
These three
qualitatively distinct regimes are shown in Fig.~\ref{regimes-graph}.

%***HERE WE PUT THAT PLOT THAT WILL CLARIFY EVERYTHING****

%%%%%%%%%%%%%%%%%%%%%%%%%%%%%%%%%

\subsection{A comment about fine-tuning}

The emergence of an energy window over
which the theory seems to walk is controlled by the choice $\alpha \ll 1$.
This might be seen as a signal of fine-tuning, since this choice is what
makes the $d\sim 4$ VEV dominate over the more relevant gaugino condensate.

In field theory,
 fine-tuned parameters are unnatural ones that are
renormalized {\it additively} (their RG equations involve
operator mixing),
and for which the experimental value
turns out to be parametrically smaller that
the natural one dictated by the additive renormalization
from other couplings in the theory.
If this is the case, setting the value of the unnatural coupling in the IR to be arbitrarily small
requires to precisely choose the UV boundary conditions of the RG flow so that
large cancellations occur in the running.
The reason why fine-tuned parameters are looked upon with suspicion is
 that their RG flow cannot be extrapolated to arbitrarily large
scales, without running into difficulties in the
interpretation of the couplings at high energies, which generically
indicates that the model should be completed in the UV.

The eq.~(\ref{approx}) automatically implies that this is not happening in 
our case.
The upper bound on $\cot\alpha$ ensures that the solution of the background equations,
and hence all the derived quantities, can be extrapolated up to arbitrary high energies without
facing singularities or pathological behaviors.
Conversely, if this condition is violated by the choice of $\alpha$ and $c$,
then in tracking back the gauge coupling from the IR up to high scales will ultimately
result into running into a singularity, that signals some sort of pathology of the RG flow.
The fact that we impose the upper limit of $\cot\alpha$ avoids such singular behavior,
and within the context of this paper allows us to disregard the possible problems
related with fine-tuning.

However, it must be emphasized that this approach is very conservative,
probably much more so that is needed in the context of dynamical electro-weak symmetry breaking, which is our original motivation.
As explained in the introduction, a technicolor model is by definition incomplete,
since by itself it does not yield a natural mechanism for generating the standard-model
fermion masses.
In any realistic model, one must assume that the technicolor dynamics is embedded in
a more general dynamical theory (ETC), and the scale at which this happens might well
be just few orders of magnitude above the electro-weak scale.
Even if the extrapolation to large scales of the background geometry towards the UV
runs into a singularity, not necessarily this implies that the solution hence constructed
has to be discarded, because the embedding into a more general theory
will anyhow modify the equations.
In constructing a realistic model, one might hence consider relaxing the bound
in eq.~(\ref{approx}), and consequently on the four-dimensional 't Hooft coupling.

%%%%%%%%%%%%%%%%%%%%%%%%%%%%%%%%%%%
%%%%%%%%%%%%%%%%%%%%%%%%%%%%%%%%%%%
%%%%%%%%%%%%%%%%%%%%%%%%%%%%%%%%%%%
%%%%%%%%%%%%%%%%%%%%%%%%%%%%%%%%%%%
%\newpage
\section{Summary and Outlook}

We reconsidered a class of Type IIB backgrounds that have been studied extensively
in the literature, found a new solution, and developed it as a
systematic expansion starting from the approximation yield by a limiting case.
We studied the set of operators (VEV's and insertions)
that it corresponds to in the dual field theory,
and the approximate symmetries of the background in the three regions
in which the background behaves differently.
The main result we obtained is that an appropriately defined four-dimensional gauge coupling
exhibits the features of a walking theory.
This coupling can be rewritten as a function of the renormalization scale as
\beqs
\lambda&=&\frac{g^2N_c}{8\pi^2}\,\simeq\,
\frac{\lambda^{\ast}\coth\r}{\left(1+2e^{-4\r_*}\left(
\sinh(4\r)-4\r\right)\right)^{1/3}},
\eeqs
where the constants $\lambda^{\ast}$ and $\r_*$ can be written in terms of the
parameters $\alpha$ and $c$ defined in the body of the paper as
\beqs
\lambda^{\ast}&=&\frac{N_c}{c\cos\alpha}\,,\\
\r_*&=&\frac{1}{4}\ln\left(2\cot^3\alpha\right)\,=\,\frac{3}{2}\ln\frac{\Lambda_*-\Lambda_0+\Lambda}{\Lambda}\,,
\eeqs
where $\r$ is related to the renormalization scale $\mu$ as
\beqs
\nonumber
\rho&=&\frac{3}{2}\ln\frac{\mu-\Lambda_0+\Lambda}{\Lambda}\,,
\eeqs
with $\Lambda_0$ the confinement scale, $\Lambda$ the spontaneous symmetry
breaking scale, and $\Lambda_*$ the walking scale,
and where the upper bound on the 't Hooft coupling
\beqs
\lambda^{\ast}&\lsim&\frac{1}{2\r_*}\,,
\eeqs
ensures that $\l$ is well defined all the way to $\r\to0$.

We showed that for $\mu<\Lambda$ the isometries of the internal manifold in the background, and  the
R-symmetries of the dual theory, are spontaneously broken by the VEV of a dimension-3 operator.
We showed that there is an approximate dilatation symmetry
for $\mu\gg \Lambda_*$.
In the intermediate region the dynamics is dominated by the VEV of a marginally irrelevant operator, while in the
far UV by the insertion of a dimension-6 operator.
It would be nice to make a dedicated study of the dual field theory in the
intermediate region.

All of the above agrees with the idea that this is a model in which the
dynamics is walking.
What is next?
There are at least four directions in which to further develop this study,
all of which might yield very interesting and useful results.

The first and most urgent question concerns the spectrum of the theory.
As anticipated, a very important open question in
the context of walking technicolor has to do with the presence of
light scalar and pseudo-scalar degrees of freedom, in connection with
the spontaneous breaking of internal and dilatational approximate symmetries.
Studying the spectrum of fluctuations in the background should shed some light on this
problem, allowing to compute the masses of possible light degrees of freedom,
and in particular allowing to identify the parametric dependence on the fundamental scales and
couplings present in the background.

A second important, though more challenging, study
should lead to the identification of the field theory dual to the present background.
Both the field theory in the UV, and the spectrum of anomalous dimensions in the intermediate and
IR regions are of utmost interest for model-building, as stressed in the introductory sections.

Third, the study of gauge invariant observables in this background, like Wilson and 't Hooft loops, or Domain Walls, may give
further information on the dual QFT.

In the fourth place, this preliminary study opens the way to a large possible set of investigations in a
broad class of models. The fact that we discovered a class of solutions yielding a
background which has some of the features of a walking theory, in a model that had already been
extensively analyzed in the literature, suggests that possibly large numbers of string-motivated
set-ups might lead to analogous results. It would be very interesting to understand how general the
occurrence of the dynamical properties discussed here is. It might even be that much simpler set-ups
admit solutions that {\it walk} over some energy interval.
The powerful computational techniques of the string-gauge theory duality
represent a wonderful opportunity for model building, and for better understanding the mechanism leading to electro-weak symmetry breaking. A more extensive survey of
the models in which this possibility is realized would be very interesting.

Finally, the original motivation of all of this is electro-weak symmetry breaking,
but the model we studied does not have any electro-weak symmetry to start.
It would be very interesting to understand how to effectively couple backgrounds
as the one described here to a weakly-coupled sector with the gauge symmetries
of the Standard Model and construct a complete realization of a dynamical
mechanism for electro-weak symmetry breaking with walking behavior. A possible way to go would be to add the `Standard Model' as
probe branes in this background (the embedding should not preserve any SUSY). In this case, the Standard Model dynamics
would feel the influence of the Technicolor sector, but not the other way around. In this `quenched' approach,
we could apply all the technology
developed in recent years, see \cite{Erdmenger:2007cm} for a review.

%%%%%%%%%%%%%%%%%%%%%%%%%%%%%%%%%%%%%%%%%%%%%%%%%%%%%%%%%%%%%%%%%%%%%%
%% Acknowledgments %%%%%%%%%%%%%%%%%%%%%%%%%%%%%%%%%%%%%%%%%%%%%%%%%%%
%%%%%%%%%%%%%%%%%%%%%%%%%%%%%%%%%%%%%%%%%%%%%%%%%%%%%%%%%%%%%%%%%%%%%%
\vspace{1.0cm}
\begin{acknowledgments}
The work of MP is supported in part  by the Wales Institute of
Mathematical and Computational Sciences. Carlos Nunez thanks Nick Evans, 
Angel Paredes and Ed Threlfall for discussions on a related topic.

\end{acknowledgments}

%%%%%%%%%%%%%%%%%%%%%%%%%%%%%%%%%%%%%%%%%%%%%%%%%%%%%%%%%%%%%%%%%%%%%%
%%%  Bibliography  %%%%%%%%%%%%%%%%%%%%%%%%%%%%%%%%%%%%%%%%%%%%%%%%%%%
%%%%%%%%%%%%%%%%%%%%%%%%%%%%%%%%%%%%%%%%%%%%%%%%%%%%%%%%%%%%%%%%%%%%%%


\begin{thebibliography}{99}


\bibitem{TC}
 S.~Weinberg,
  %``Implications Of Dynamical Symmetry Breaking: An Addendum,''
  Phys.\ Rev.\ D {\bf 19}, 1277 (1979);
  %%CITATION = PHRVA,D19,1277;%%
L.~Susskind,
  %``Dynamics Of Spontaneous Symmetry Breaking In The Weinberg-Salam Theory,''
  Phys.\ Rev.\ D {\bf 20}, 2619 (1979);
  %%CITATION = PHRVA,D20,2619;%%
 S.~Weinberg,
  %``Implications Of Dynamical Symmetry Breaking,''
  Phys.\ Rev.\ D {\bf 13}, 974 (1976).
  %%CITATION = PHRVA,D13,974;%%




\bibitem{EWCL}
 T.~Appelquist and C.~W.~Bernard,
  %``Strongly Interacting Higgs Bosons,''
  Phys.\ Rev.\  D {\bf 22}, 200 (1980);
  %%CITATION = PHRVA,D22,200;%%
 A.~C.~Longhitano,
  %``Heavy Higgs Bosons In The Weinberg-Salam Model,''
  Phys.\ Rev.\  D {\bf 22}, 1166 (1980);
  %%CITATION = PHRVA,D22,1166;%%
% A.~C.~Longhitano,
  %``Low-Energy Impact Of A Heavy Higgs Boson Sector,''
  Nucl.\ Phys.\  B {\bf 188}, 118 (1981);
  %%CITATION = NUPHA,B188,118;%%
 T.~Appelquist and G.~H.~Wu,
  %``The Electroweak chiral Lagrangian and new precision measurements,''
  Phys.\ Rev.\  D {\bf 48}, 3235 (1993)
  [arXiv:hep-ph/9304240];
  %%CITATION = PHRVA,D48,3235;%%
% T.~Appelquist and G.~H.~Wu,
  %``The Electroweak Chiral Lagrangian And CP Violating Effects In Technicolor
  %Theories,''
  Phys.\ Rev.\  D {\bf 51}, 240 (1995)
  [arXiv:hep-ph/9406416].
  %%CITATION = PHRVA,D51,240;%%



\bibitem{PT}
 M.~E.~Peskin and T.~Takeuchi,
  %``Estimation of oblique electroweak corrections,''
  Phys.\ Rev.\ D {\bf 46}, 381 (1992).
  %%CITATION = PHRVA,D46,381;%%

\bibitem{NDA}
See for instance  H.~Georgi,
  %``Generalized Dimensional Analysis,''
  Phys.\ Lett.\  B {\bf 298}, 187 (1993)
  [arXiv:hep-ph/9207278],
  %%CITATION = PHLTA,B298,187;%%
and references therein.



  \bibitem{Barbieri}
R.~Barbieri, A.~Pomarol, R.~Rattazzi and A.~Strumia,
  %``Electroweak symmetry breaking after LEP1 and LEP2,''
  Nucl.\ Phys.\ B {\bf 703}, 127 (2004)
  [arXiv:hep-ph/0405040].
  %%CITATION = HEP-PH 0405040;%%

\bibitem{ETC}
 S.~Dimopoulos and L.~Susskind,
  %``Mass Without Scalars,''
  Nucl.\ Phys.\ B {\bf 155}, 237 (1979);
  %%CITATION = NUPHA,B155,237;%%
 E.~Eichten and K.~D.~Lane,
  %``Dynamical Breaking Of Weak Interaction Symmetries,''
  Phys.\ Lett.\ B {\bf 90}, 125 (1980).
  %%CITATION = PHLTA,B90,125;%%








\bibitem{WTC}
 B.~Holdom,
  %``Techniodor,''
  Phys.\ Lett.\ B {\bf 150}, 301 (1985);
  %%CITATION = PHLTA,B150,301;%%
    K.~Yamawaki, M.~Bando and K.~i.~Matumoto,
  %``Scale Invariant Technicolor Model And A Technidilaton,''
  Phys.\ Rev.\ Lett.\  {\bf 56}, 1335 (1986);
  %%CITATION = PRLTA,56,1335;%%
T.~W.~Appelquist, D.~Karabali and L.~C.~R.~Wijewardhana,
  %``Chiral Hierarchies And The Flavor Changing Neutral Current Problem In
  %Technicolor,''
  Phys.\ Rev.\ Lett.\  {\bf 57}, 957 (1986).
  %%CITATION = PRLTA,57,957;%%

  \bibitem{seiberg}
  N.~Seiberg,
  %``Exact Results On The Space Of Vacua Of Four-Dimensional Susy Gauge
  %Theories,''
  Phys.\ Rev.\  D {\bf 49}, 6857 (1994)
  [arXiv:hep-th/9402044];
  %%CITATION = PHRVA,D49,6857;%%
  N.~Seiberg,
  %``Electric - magnetic duality in supersymmetric nonAbelian gauge theories,''
  Nucl.\ Phys.\  B {\bf 435}, 129 (1995)
  [arXiv:hep-th/9411149];
  %%CITATION = NUPHA,B435,129;%%
  K.~A.~Intriligator and N.~Seiberg,
  %``Lectures on supersymmetric gauge theories and electric-magnetic  duality,''
  Nucl.\ Phys.\ Proc.\ Suppl.\  {\bf 45BC}, 1 (1996)
  [arXiv:hep-th/9509066].
  %%CITATION = NUPHZ,45BC,1;%%

  \bibitem{cascade}
A useful summary of related results can be found in
M.~J.~Strassler,
  %``The duality cascade,''
  arXiv:hep-th/0505153.
  %%CITATION = HEP-TH/0505153;%%


  \bibitem{CC}
 A.~Manohar and H.~Georgi,
  %``Chiral Quarks And The Nonrelativistic Quark Model,''
  Nucl.\ Phys.\  B {\bf 234}, 189 (1984).
  %%CITATION = NUPHA,B234,189;%%



\bibitem{lattice}
F.~Karsch and M.~Lutgemeier,
  %``Deconfinement and chiral symmetry restoration in an SU(3) gauge theory
  %with adjoint fermions,''
  Nucl.\ Phys.\  B {\bf 550}, 449 (1999)
  [arXiv:hep-lat/9812023].
  %%CITATION = NUPHA,B550,449;%%


\bibitem{SS}
 T.~Sakai and S.~Sugimoto,
  %``Low energy hadron physics in holographic QCD,''
  Prog.\ Theor.\ Phys.\  {\bf 113}, 843 (2005)
  [arXiv:hep-th/0412141];
  %%CITATION = PTPKA,113,843;%%
  O.~Aharony, J.~Sonnenschein and S.~Yankielowicz,
  %``A holographic model of deconfinement and chiral symmetry restoration,''
  Annals Phys.\  {\bf 322}, 1420 (2007)
  [arXiv:hep-th/0604161].
  %%CITATION = APNYA,322,1420;%%




\bibitem{AS}
  T.~Appelquist and F.~Sannino,
  %``The physical spectrum of conformal SU(N) gauge theories,''
  Phys.\ Rev.\ D {\bf 59}, 067702 (1999)
  [arXiv:hep-ph/9806409].
  %%CITATION = HEP-PH 9806409;%%

\bibitem{AdS/TC}
J.~Hirn and V.~Sanz,
  %``A negative S parameter from holographic technicolor,''
  Phys.\ Rev.\ Lett.\  {\bf 97}, 121803 (2006)
  [arXiv:hep-ph/0606086],
  %%CITATION = PRLTA,97,121803;%%
%J.~Hirn and V.~Sanz,
  %``The fifth dimension as an analogue computer for strong interactions at the
  %LHC,''
  JHEP {\bf 0703}, 100 (2007)
  [arXiv:hep-ph/0612239];
  %%CITATION = JHEPA,0703,100;%%
   D.~K.~Hong and H.~U.~Yee,
  %``Holographic estimate of oblique corrections for technicolor,''
  Phys.\ Rev.\  D {\bf 74}, 015011 (2006)
  [arXiv:hep-ph/0602177];
  %%CITATION = PHRVA,D74,015011;%%
 M.~Piai,
 %   ``Precision electro-weak parameters from AdS(5), localized kinetic terms and
  %anomalous dimensions,''
  arXiv:hep-ph/0608241,
  %%CITATION = HEP-PH 0608241;%%
 % M.~Piai,
  %``Walking in the third millennium,''
  arXiv:hep-ph/0609104,
  %%CITATION = HEP-PH/0609104;%%
  %``Vector mesons from AdS/TC to the LHC,''
  arXiv:0704.2205 [hep-ph];
  %%CITATION = ARXIV:0704.2205;%%
     C.~D.~Carone, J.~Erlich and J.~A.~Tan,
  %``Holographic bosonic technicolor,''
  arXiv:hep-ph/0612242;
  %%CITATION = HEP-PH/0612242;%%
M~Fabbrichesi, M.~Piai, L.~Vecchi
arXiv:0804.0124 [hep-ph];
  %%CITATION = ARXIV:0804.0124;%%
    K.~Haba, S.~Matsuzaki and K.~Yamawaki,
  %``$S$ Parameter in the Holographic Walking/Conformal Technicolor,''
  arXiv:0804.3668 [hep-ph];
  %%CITATION = ARXIV:0804.3668;%%
  J.~Hirn, A.~Martin and V.~Sanz,
  %``Describing viable technicolor scenarios,''
  arXiv:0807.2465 [hep-ph];
  %%CITATION = ARXIV:0807.2465;%%
 D.~D.~Dietrich and C.~Kouvaris,
  %``Generalised bottom-up holography and walking technicolour,''
  arXiv:0809.1324 [hep-ph].
  %%CITATION = ARXIV:0809.1324;%%

\bibitem{APS} One example of such a construction has been studied in
 T.~Appelquist, M.~Piai and R.~Shrock,
  %``Fermion masses and mixing in extended technicolor models,''
  Phys.\ Rev.\ D {\bf 69}, 015002 (2004)
  [arXiv:hep-ph/0308061];
  %%CITATION = HEP-PH 0308061;%%
  %T.~Appelquist, M.~Piai and R.~Shrock,
  %``Lepton dipole moments in extended technicolor models,''
  Phys.\ Lett.\ B {\bf 593}, 175 (2004)
  [arXiv:hep-ph/0401114];
  %%CITATION = HEP-PH 0401114;%%
   %T.~Appelquist, M.~Piai and R.~Shrock,
  %``Quark dipole operators in extended technicolor models,''
  Phys.\ Lett.\ B {\bf 595}, 442 (2004)
  [arXiv:hep-ph/0406032];
  %%CITATION = HEP-PH 0406032;%%
  T.~Appelquist, N.~D.~Christensen, M.~Piai and R.~Shrock,
  %``Flavor-changing processes in extended technicolor,''
  Phys.\ Rev.\ D {\bf 70}, 093010 (2004)
  [arXiv:hep-ph/0409035].
  %%CITATION = HEP-PH 0409035;%%

\bibitem{dilaton}
  W.~A.~Bardeen, C.~N.~Leung and S.~T.~Love,
  %``The Dilaton And Chiral Symmetry Breaking,''
  Phys.\ Rev.\ Lett.\  {\bf 56}, 1230 (1986);
  %%CITATION = PRLTA,56,1230;%%
    M.~Bando, K.~i.~Matumoto and K.~Yamawaki,
  %``TECHNIDILATON,''
  Phys.\ Lett.\  B {\bf 178}, 308 (1986);
  %%CITATION = PHLTA,B178,308;%%
    B.~Holdom and J.~Terning,
  %``A Light Dilaton In Gauge Theories?,''
  Phys.\ Lett.\  B {\bf 187}, 357 (1987),
  %%CITATION = PHLTA,B187,357;%%
 %   B.~Holdom and J.~Terning,
  %``NO LIGHT DILATON IN GAUGE THEORIES,''
  Phys.\ Lett.\  B {\bf 200}, 338 (1988);
  %%CITATION = PHLTA,B200,338;%%
W.~D.~Goldberger, B.~Grinstein and W.~Skiba,
  %``Light scalar at LHC: the Higgs or the dilaton?,''
  Phys.\ Rev.\ Lett.\  {\bf 100}, 111802 (2008)
  [arXiv:0708.1463 [hep-ph]].
  %%CITATION = PRLTA,100,111802;%%


\bibitem{BZ}
 T.~Banks and A.~Zaks,
  %``On The Phase Structure Of Vector-Like Gauge Theories With Massless
  %Fermions,''
  Nucl.\ Phys.\  B {\bf 196}, 189 (1982).
  %%CITATION = NUPHA,B196,189;%%
 
%\bibitem{seiberg}

\bibitem{KS}
See for instance   M.~Kurachi and R.~Shrock,
  %``Study of the change from walking to non-walking behavior in a vectorial
  %gauge theory as a function of N(f),''
  JHEP {\bf 0612}, 034 (2006)
  [arXiv:hep-ph/0605290],
  %%CITATION = JHEPA,0612,034;%%
and references therein.

\bibitem{NSVZ}
V.~A.~Novikov, M.~A.~Shifman, A.~I.~Vainshtein and V.~I.~Zakharov,
  %``Exact Gell-Mann-Low Function Of Supersymmetric Yang-Mills Theories From
  %Instanton Calculus,''
  Nucl.\ Phys.\  B {\bf 229}, 381 (1983);
  %%CITATION = NUPHA,B229,381;%%
  M.~A.~Shifman and A.~I.~Vainshtein,
  %``Solution of the Anomaly Puzzle in SUSY Gauge Theories and the Wilson
  %Operator Expansion,''
  Nucl.\ Phys.\  B {\bf 277}, 456 (1986)
  [Sov.\ Phys.\ JETP {\bf 64}, 428 (1986\ ZETFA,91,723-744.1986)].
  %%CITATION = ZETFA,91,723;%%

\bibitem{exactbeta}
 T.~A.~Ryttov and F.~Sannino,
  %``Supersymmetry Inspired QCD Beta Function,''
  Phys.\ Rev.\  D {\bf 78}, 065001 (2008)
  [arXiv:0711.3745 [hep-th]].
  %%CITATION = PHRVA,D78,065001;%%
 
\bibitem{AF}
 T.~Appelquist, G.~T.~Fleming and E.~T.~Neil,
  %``Lattice Study of the Conformal Window in QCD-like Theories,''
  Phys.\ Rev.\ Lett.\  {\bf 100}, 171607 (2008)
  [arXiv:0712.0609 [hep-ph]].
  %%CITATION = PRLTA,100,171607;%%
 
\bibitem{DSS}
T.~DeGrand, Y.~Shamir and B.~Svetitsky,
  %``Zero of the discrete beta function in SU(3) lattice gauge theory with color
  %sextet fermions,''
  Phys.\ Rev.\  D {\bf 78}, 031502 (2008)
  [arXiv:0803.1707 [hep-lat]];
  %%CITATION = PHRVA,D78,031502;%%
%T.~DeGrand, Y.~Shamir and B.~Svetitsky,
  %``Phase structure of SU(3) gauge theory with two flavors of
  %symmetric-representation fermions,''
  arXiv:0812.1427 [hep-lat].
  %%CITATION = ARXIV:0812.1427;%%

\bibitem{walkthelattice}
S.~Catterall and F.~Sannino,
  %``Minimal walking on the lattice,''
  Phys.\ Rev.\  D {\bf 76}, 034504 (2007)
  [arXiv:0705.1664 [hep-lat]];
  %%CITATION = PHRVA,D76,034504;%%
S.~Catterall, J.~Giedt, F.~Sannino and J.~Schneible,
  %``Phase diagram of SU(2) with 2 flavors of dynamical adjoint quarks,''
  JHEP {\bf 0811}, 009 (2008)
  [arXiv:0807.0792 [hep-lat]];
  %%CITATION = JHEPA,0811,009;%%
  A.~Deuzeman, M.~P.~Lombardo and E.~Pallante,
  %``The physics of eight flavours,''
  Phys.\ Lett.\  B {\bf 670}, 41 (2008)
  [arXiv:0804.2905 [hep-lat]];
  %%CITATION = PHLTA,B670,41;%%
 L.~Del Debbio, A.~Patella and C.~Pica,
  %``Higher representations on the lattice: numerical simulations. SU(2) with
  %adjoint fermions,''
  arXiv:0805.2058 [hep-lat];
  %%CITATION = ARXIV:0805.2058;%%
  A.~J.~Hietanen, J.~Rantaharju, K.~Rummukainen and K.~Tuominen,
  %``Spectrum of SU(2) lattice gauge theory with two adjoint Dirac flavours,''
  arXiv:0812.1467 [hep-lat].
  %%CITATION = ARXIV:0812.1467;%%

%%%%%%%%%%%%%%%%%%%%%%%%%%%%%%%%%%%%%%%
%%%%%%%%%%%%%%%%%%%%%%%%%%%%%%%%%%%%%%%
%%%%%%%%%%%%%%%%%%%%%%%%%%%%%%%%%%%%%%%%


%\cite{Itzhaki:1998dd}
\bibitem{Itzhaki:1998dd}
  N.~Itzhaki, J.~M.~Maldacena, J.~Sonnenschein and S.~Yankielowicz,
  %``Supergravity and the large N limit of theories with sixteen
  %supercharges,''
  Phys.\ Rev.\  D {\bf 58}, 046004 (1998)
  [arXiv:hep-th/9802042].
  %%CITATION = PHRVA,D58,046004;%%


%\cite{Maldacena:2000yy}
\bibitem{Maldacena:2000yy}
  J.~M.~Maldacena and C.~Nunez,
  %``Towards the large N limit of pure N = 1 super Yang Mills,''
  Phys.\ Rev.\ Lett.\  {\bf 86}, 588 (2001)
  [arXiv:hep-th/0008001].
  %%CITATION = PRLTA,86,588;%%


%\cite{Andrews:2006aw}
\bibitem{Andrews:2006aw}
  R.~P.~Andrews and N.~Dorey,
  %``Deconstruction of the Maldacena-Nunez compactification,''
  Nucl.\ Phys.\  B {\bf 751}, 304 (2006)
  [arXiv:hep-th/0601098].
  %%CITATION = NUPHA,B751,304;%%


%\cite{Bershadsky:1995qy}
\bibitem{Bershadsky:1995qy}
  M.~Bershadsky, C.~Vafa and V.~Sadov,
  %``D-Branes and Topological Field Theories,''
  Nucl.\ Phys.\  B {\bf 463}, 420 (1996)
  [arXiv:hep-th/9511222].
  %%CITATION = NUPHA,B463,420;%%



%\cite{Casero:2006pt}
\bibitem{Casero:2006pt}
  R.~Casero, C.~Nunez and A.~Paredes,
  %``Towards the string dual of N = 1 SQCD-like theories,''
  Phys.\ Rev.\  D {\bf 73}, 086005 (2006)
  [arXiv:hep-th/0602027].
  %%CITATION = PHRVA,D73,086005;%%





%\cite{HoyosBadajoz:2008fw}
\bibitem{HoyosBadajoz:2008fw}
  C.~Hoyos-Badajoz, C.~Nunez and I.~Papadimitriou,
  %``Comments on the String dual to N=1 SQCD,''
  Phys.\ Rev.\  D {\bf 78}, 086005 (2008)
  [arXiv:0807.3039 [hep-th]].
  %%CITATION = PHRVA,D78,086005;%%
See also
%\cite{Casero:2007jj}
%\bibitem{Casero:2007jj}
  R.~Casero, C.~Nunez and A.~Paredes,
  %``Elaborations on the String Dual to N=1 SQCD,''
  Phys.\ Rev.\  D {\bf 77}, 046003 (2008)
  [arXiv:0709.3421 [hep-th]].
  %%CITATION = PHRVA,D77,046003;%%


%\cite{Witten:1994ev}
\bibitem{Witten:1994ev}
  E.~Witten,
  %``Supersymmetric Yang-Mills theory on a four manifold,''
  J.\ Math.\ Phys.\  {\bf 35}, 5101 (1994)
  [arXiv:hep-th/9403195].
  %%CITATION = JMAPA,35,5101;%%




%\cite{DiVecchia:2002ks}
\bibitem{DiVecchia:2002ks}
  P.~Di Vecchia, A.~Lerda and P.~Merlatti,
  %``N = 1 and N = 2 super Yang-Mills theories from wrapped branes,''
  Nucl.\ Phys.\  B {\bf 646}, 43 (2002)
  [arXiv:hep-th/0205204].
  %%CITATION = NUPHA,B646,43;%%
%\cite{Bertolini:2002yr}
%\bibitem{Bertolini:2002yr}
  M.~Bertolini and P.~Merlatti,
  %``A note on the dual of N = 1 super Yang-Mills theory,''
  Phys.\ Lett.\  B {\bf 556}, 80 (2003)
  [arXiv:hep-th/0211142].
  %%CITATION = PHLTA,B556,80;%%



%\cite{Nunez:2003cf}
\bibitem{Nunez:2003cf}
  C.~Nunez, A.~Paredes and A.~V.~Ramallo,
  %``Flavoring the gravity dual of N = 1 Yang-Mills with probes,''
  JHEP {\bf 0312}, 024 (2003)
  [arXiv:hep-th/0311201].
  %%CITATION = JHEPA,0312,024;%%

%\cite{Gubser:2000nd}
\bibitem{Gubser:2000nd}
  S.~S.~Gubser,
  %``Curvature singularities: The good, the bad, and the naked,''
  Adv.\ Theor.\ Math.\ Phys.\  {\bf 4}, 679 (2000)
  [arXiv:hep-th/0002160].
  %%CITATION = 00203,4,679;%%

%\cite{Gursoy:2003hf}
\bibitem{Gursoy:2003hf}
  U.~Gursoy, S.~A.~Hartnoll and R.~Portugues,
  %``The chiral anomaly from M theory,''
  Phys.\ Rev.\  D {\bf 69}, 086003 (2004)
  [arXiv:hep-th/0311088].
  %%CITATION = PHRVA,D69,086003;%%


%\cite{Maldacena:2000mw}
\bibitem{Maldacena:2000mw}
  J.~M.~Maldacena and C.~Nunez,
  %``Supergravity description of field theories on curved manifolds and a
%no  go
  %theorem,''
  Int.\ J.\ Mod.\ Phys.\  A {\bf 16}, 822 (2001)
  [arXiv:hep-th/0007018].
  %%CITATION = IMPAE,A16,822;%%

%\cite{Kazakov:2002jd}
\bibitem{Kazakov:2002jd}
  D.~I.~Kazakov,
  %``Ultraviolet fixed points in gauge and SUSY field theories in extra
  %dimensions,''
  JHEP {\bf 0303}, 020 (2003)
  [arXiv:hep-th/0209100].
  %%CITATION = JHEPA,0303,020;%%

%\cite{Erdmenger:2007cm}
\bibitem{Erdmenger:2007cm}
  J.~Erdmenger, N.~Evans, I.~Kirsch and E.~Threlfall,
  %``Mesons in Gauge/Gravity Duals - A Review,''
  Eur.\ Phys.\ J.\  A {\bf 35}, 81 (2008)
  [arXiv:0711.4467 [hep-th]].
  %%CITATION = EPHJA,A35,81;%%


\end{thebibliography}
\end{document}